\documentstyle[preprint,aps]{revtex}
\begin{document}
\draft

\title{
\begin{flushright}
{\rm MSUCL-946}
\end{flushright}
Effects of Compression and Collective Expansion\\
on Particle Emission from Central Heavy-Ion Reactions
}
\author{P. Danielewicz\thanks{e-mail:
danielewicz@nscl.nscl.msu.edu} }
\address{
National Superconducting Cyclotron Laboratory and\\
Department of Physics and Astronomy, Michigan State University,
\\
East Lansing, Michigan 48824, USA\\
}
\date{\today}
\maketitle

\begin{abstract}
Conditions under which compression occurs and collective
expansion develops in energetic
symmetric reactions of heavy nuclei, are analyzed, together
with their effects on emitted light baryons and pions.
Within
transport simulations, it~is shown that shock fronts
perpendicular to beam axis form in head-on reactions.
The fronts
separate hot compressed matter from normal and
propagate into projectile and target.
As impact parameter increases, the angle of inclination of
the fronts relative to beam axis decreases,
and in-between the
fronts a~weak tangential discontinuity develops.
Hot matter exposed to the vacuum in directions perpendicular to
shock motion (and parallel to fronts), starts to expand
sideways,
early within reactions.  Expansion in the direction of shock
motion follows after
the shocks propagate through nuclei, but due to the delay
does not acquire same strength.  Expansion
affects
angular distributions, mean-energy components, shapes of
spectra and mean energies of different particles emitted into
any~one direction, and further particle yields.  Both the
expansion
and a~collective motion associated with the weak discontinuity,
affect the
magnitude of sideward flow within reaction plane.  Differences
in mean particle energy components in and out of the reaction
plane in semicentral collisions, depend
sensitively
on the relative magnitude of shock speed in normal matter and
speed of
sound in hot matter.
The missing energy, considered in the past in
association with low pion-multiplicity in central reactions,
may be identified with the energy of collective expansion.
Relations are established which approximately govern the
behavior of density and entropy in the
compressed region in reactions with beam energy and impact
parameter.
\end{abstract}


\narrowtext
\section{INTRODUCTION}

Within the past two decades a considerable effort in the area
of energetic heavy-ion reactions was put into
identification of signatures of collective phenomena
in excited nuclear matter.  Particular effort was
devoted to the effects of collective
motion.  The latter has been understood as a motion
characterized
by the correlation between particle
positions and momenta, of a dynamic origin.  Early attempts
to identify signatures of collective phenomena
in single-particle observables were, generally, plagued by
ambiguities.
Thus it was anticipated~\cite{cha73,sch74,sob75} that nuclear
matter gets stopped in the
very central collisions, compressed and heated, and
subsequently expands primarily in the directions perpendicular
to the beam axis.
The~expansion was suggested~\cite{sie79} to be the
cause of
observed differences in the transverse slopes of the inclusive
pion and proton spectra.  However, it was further realized
that
the production of pions in the reactions mainly proceeds
through the $\Delta$ resonance excitation and decay, and
thus the pion slopes could reflect the
$\Delta$-decay kinematics~\cite{bro84}.
A~sideward peaking of the proton angular distributions in an
asymmetric
reaction gave a~hint of the expected
collective behavior~\cite{sto80}.  As eventually, though, the
neutron distributions had shown no such peaking~\cite{mad86},
one was led to
the conclusion that the peaking of the proton distributions was
a Coulomb effect~\cite{sch82}.
Stock~{\em et al.}~\cite{sto82} noted
that the measured pion multiplicities in central
collisions, lower than predicted within
the cascade or fireball models, could be explained by assuming
that a part of the available energy was used up for
compressing nuclear matter.  Subsequent dynamic
calculations
of collisions~\cite{ber84} have demonstrated, though, a weak
sensitivity
of the pion yields to the compressibility of matter.

Analyses of the $4\pi$ data brought fairly unambiguous
evidence~\cite{gus84} of what appeared to be a different type of
collective motion than first envisioned
in the head-on collisions.  This was a sideward deflection in
the reaction plane of the fragments moving forward and backward
in the center of mass in semicentral collisions.  The
deflection was quantified
with a flow angle or with an average momentum in the reaction
plane
at a given rapidity.  At~midrapidity, at~high beam energies,
a~preference was found for
fragments to be emitted out of the reaction
plane~\cite{dan87,gut89}.  It~has not been obvious whether this
preference
was of a~dynamic origin or due to shadowing.    At~low
energies the
emission in the reaction plane was found to be
enhanced~\cite{wil90}.

Recent years brought new important results
on the motion in central reactions.  Specifically,
Schussler~{\em et al.}~\cite{bar91,bau93} have investigated
the mean c.m. energies
of fragments emitted from the central reactions with Ag in
emulsion.  For~the
incident $^{36}$Ar ions at $65 \pm 15$ MeV, and the incident
$^{94}$Kr ions  at $95 \pm 25$ MeV,
they observed a rise of these energies with fragment
charge, faster
than anticipated on the basis of Coulomb considerations alone.
For~the two reactions the rise was
consistent with a collective expansion characterized by
the energy of $\sim$3 MeV/nucleon and
$\raisebox{-.5ex}{$\stackrel{>}{\scriptstyle\sim}$ }$7
MeV/nucleon, respectively.
No rapid rise
was observed for the incident
$^{16}$O ions at $210\pm 10$ MeV.  The FOPI collaboration at
GSI
has further analysed the spectra and mean energies
of fragments with different charge emitted from central Au~+~Au
collisions at
beam energies in the range
(150-400)~MeV/nucleon~\cite{rei93,jeo94}.  They have found that
the spectra of
fragments with large $Z$ were too broad and, respectively, the
mean energies too high to be accounted for by a model of purely
statistical decay.  Within preliminary analysis~\cite{rei93},
data description improved once an
assumption of the collective expansion at the time of decay was
introduced, of an energy $\sim$18 MeV/nucleon at the beam
energy
150 MeV/nucleon, and of an energy $\sim$51 MeV/nucleon at 400
MeV/nucleon,
respectively.  In~a~later more cautious
investigation~\cite{jeo94}, the collective energy was bounded from
below with $\sim 10$ MeV/nucleon in the~150~MeV/nucleon
reaction.
The fragment yields were
separately analysed to extract entropy~\cite{kuh93} within
the so-called Quantum Statistical Model~(QSM).  Low entropy
values
were extracted consistent with low temperatures at
reasonable values of
freeze-out density.  (These values were lower than those
extracted in the past from light-fragment
yields~\cite{dos88a}).
The collective expansion could alleviate
but not eliminate discrepancies between the temperatures and
energy available in the center of mass.
Further analysis of spectra for collective motion in a~central
100~MeV/nucleon Au~+~Au reaction was done~in~\cite{hsi94}.

On the theoretical side, emission from central collisions has
been analysed within a~transport model with a dynamic
light-fragment production ($A\le 3$)~\cite{dan92}.  Within this
model the compression and excitation of nuclear matter in
central collisions
was observed, followed by an expansion.  The mean energies of
emitted fragments
were observed to rise with the fragment mass, and a~general
consistency
has been established between the rise with the mass and the collective
energy determined from the dynamics.  The calculated energies
partially
agreed and partially disagreed with the Plastic Ball
data~\cite{dos88}.  A degree of agreement
of the calculations was found~\cite{pog93}, on the other hand,
with the  preliminary FOPI data on light fragments.
Within the calculations~\cite{dan92}, the expansion affected,
besides energies, the fragment angular-distributions.
A~midrapidity
out-of-the-reaction-plane peaking at intermediate
impact parameters in a heavy system, in a quantitative
agreement with
the data~\cite{gut89} (see also Refs.~\cite{koc91,har93}), was
identified as a continuation, with the rise of impact
parameter, of a 90$^\circ$ peaking in the polar angle in
head-on reactions.

Spurred by the recent experimental
evidence~\cite{bar91,bau93,rei93,jeo94,hsi94}
and some forthcoming results~\cite{lac92,pog93,rai93}, this
study extends that carried out in~\cite{dan92}.
Very preliminary results have been presented~in~\cite{dan93}.
Conditions under which a~compression occurs in
energetic reactions of heavy nuclei, and
a~collective expansion develops, are analyzed in detail,
together with their effects on emitted light baryons and pions.
Relations are established which approximately govern the
behavior of density and entropy in the
compressed region in reactions, with beam energy and impact
parameter.
Anisotropies in particle emissions in the polar angle in
head-on
reactions and in the azimuthal angle in semicentral
reactions,
are explained in terms of a~different starting time for
expansions in different directions.
Besides the effects of collective expansion on
baryon and pion spectra, effects on particle yields
are demonstrated.
The effects on pions turn out to
be dramatic, contrary to naive expectations.
The~role of expansion
in the formation of hollow structures at lower
energies~\cite{mor92,bau92,bao92,hxu93} is elucidated.
The~analysis renders an~understanding of both recent
and older data.

The outline of the paper is as follows.
Section~\ref{Model} describes the model used for reaction
analysis.   Section~\ref{Central} is devoted to the dynamics
and some  observables from symmetric head-on reactions of
heavy-nuclei.  The~gained experience is next utilized in the
study of semi-central reactions~in~Sec.~\ref{Semi}.
Section~\ref{Pion} is devoted to the effects of collective
expansion on pion yields, effects of the expansion on spectra
of relativistic particles in general, and to shadowing of pions
in semi-central reactions.  Results on entropy and on collective
expansion energy from measurements and simulations are
discussed~in~Sec.~\ref{frapro}.  Section~\ref{ints} discusses
the~sensitivity of collective energy and its components to
nuclear compressibility and elementary cross sections, and
further the mechanism of formation of exotic structures at low
energies.  The~results are summarized~Sec.~\ref{Conclu}.

\section{TRANSPORT MODEL}
\label{Model}

Reaction simulations are carried out within a transport
model with explicit nucleon, deuteron,
$A = 3$ cluster, pion, and delta and $N^*$ degrees of
freedom~\cite{dan91,dan92}.  The phase-space occupation functions $f_X$
of stable particles satisfy transport equations
following
from the nonequilibrium many-body theory in the quasiparticle
limit,
\begin{eqnarray}
{\partial f_X \over \partial t} + {\partial E_X \over
\partial {\bf p}_X} {\bf \cdot} {\partial f_X \over \partial
{\bf r}} - {\partial E_X
\over \partial {\bf r}} {\bf \cdot} {\partial f_X \over
\partial
{\bf p}_X} = {\cal K}^<_X (1 \mp f_X) - {\cal K}^>_X f_X .
\label{boltz}
\end{eqnarray}
In the above $E_X$ is
the single-particle energy, and
${\cal K}^<_X$ and
${\cal K}^>_X$ are, respectively, the production and absorption
rates of the particle $X$.  The upper and lower signs in the
statistical factor on the r.h.s are for
fermions and bosons, respectively.

The deuterons within the
model are formed in the interaction of three
nucleons~\cite{dan91,gyu83,rop88}.  The formation
rate~\cite{rop88,dan91} contributing to
the deuteron production rate has a form analogous to that of
the rate associated with elastic scattering,
\begin{eqnarray}
{\cal K}^{<}_{d}({\bf P}) = &
{ 8 \over 3} & {m_d\over E_d ({\bf P})}
\sum_{N=n,p} \int
{{ d{\bf p}}\over (2\pi )^{3}} {{m_N}\over
E_{N}({\bf p})} \,
{{ d{\bf p}}_1'\over (2\pi )^{3}} {m_{N}\over
E_{N}({\bf p_1'})} \,
{{ d{\bf p}}_2'\over (2\pi )^{3}} {m_{N}\over
E_{N}({\bf p}_2')} \,
{{ d{\bf p}'}\over (2\pi )^{3}} {m_{N}\over
E_{N}({\bf
p}')}                           \cr & \times &
{1 \over 2} \overline{|{\cal M}_{ Npn\rightarrow Nd}|^2}
\,
(2\pi )^{3}\delta ({\bf P} + {\bf p} - {\bf p}_1' - {\bf p}_2'
- {\bf p}')
2\pi \delta (E_{d}({\bf P}) + E_{N}({\bf p}) \cr & - &
E_{p}({\bf p}_1') - E_{n}({\bf p}_2')-
E_{N}({\bf p}'))\,
(1 - f_{N}({\bf p}))
f_{p}({\bf p}_1')f_{n}({\bf p}_2')f_N({\bf p}')
+ \cdots ,
\label{kdeu}
\end{eqnarray}
The dots indicate other terms in the production rate.  The
factor $\overline{ |{\cal M}|^2}$ stands for the matrix element
squared summed over the final and averaged over the initial
spin directions.  A~term in the deuteron absorption rate
accounting for the deuteron breakup into nucleons has
an~analogous form to the term for the formation (\ref{kdeu}), with
statistical factors for the initial and final states
interchanged.  Due to the microscopic reversibility the
processes
of deuteron formation and breakup share the matrix elements
squared.  The averaged elements squared are then related with
the ratio of spin-degeneracies,
\begin{equation}
\overline{|{\cal M}_{Npn\rightarrow Nd}|^2} =
{3 \over 4}
\overline{|{\cal M}_{Nd\rightarrow Npn}|^2}.
\end{equation}
In principle, data on deuteron breakup could be
used in describing the formation as the differential
cross section for breakup is proportional to the matrix element
squared,
\begin{eqnarray}
d\sigma _{ Nd\rightarrow  Npn} = & & {1\over v_{1}}
{m_{N}\over E_N({\bf p}_{1})}
\overline{|{\cal M}_{{ Nd}\rightarrow Npn}|^2}
\, 2\pi \delta (m_d +
E_N({\bf
p}_{1}) - E_N({\bf p}_1') - E_p({\bf p}_2' ) -
E_n({\bf p}_3'))\cr & & \times (2\pi
)^{3}\delta ({\bf p}_{1} - {\bf p}_1'- {\bf p}_2' - {\bf
p}_3') \, {m_N\over E_{N}({\bf
p}_1')} {d{\bf p}_1' \over (2\pi )^{3}} \, {m_p\over
E_p({\bf p}_2' )} {d{\bf p}_2' \over (2\pi
)^{3}}  \, {m_n\over E_n({\bf p}_3')} {d{\bf
p}_3' \over (2\pi )^{3}} .
\end{eqnarray}
In practise, due to the dimensionality of three-body space, the
differential cross section is known only within a limited
range.
In the calculations, a modified impulse
approximation is thus used
\begin{eqnarray}
\protect\overline{|{\cal M}_{ {pd} \rightarrow ppn }|^2}\simeq &&F \Big\{
\left| \left< p_1' | \phi_d \right>
\right|^2 \overline{|{\cal M}_{{pn}\rightarrow{pn}}|^2} +
\left| \left< p_2' | \phi_d \right>
\right|^2 \overline{|{\cal M}_{{pn}\rightarrow{pn}}|^2} \cr && +
\left| \left< p_3' | \phi_d \right>
\right|^2 \overline{|{\cal M}_{ pp\rightarrow pp}|^2} \Big\} ,
\label{modimp}
\end{eqnarray}
where $\phi_d$ is the deuteron wave-function in
momentum space, normalized so that
\begin{eqnarray} \nonumber
\int {d{\bf p} \over  (2\pi)^3}|<p|\phi>|^2 = 1.
\end{eqnarray}
  The three terms in (\ref{modimp})
correspond to the three different nucleons being spectators.
The~$NN$~matrix elements in are proportional to the $NN$ cross
sections.  Finally, the overall normalization factor $F$ in
(\ref{modimp}) is adjusted so that the measured total deuteron
breakup cross section is reproduced as a function of bombarding
energy~\cite{dan91}.

Tritons and helions in the transport model are produced in
the interactions involving four nucleons~\cite{dan92a,dan92}.
The  production of a composite is generally suppressed if the
average nucleon occupation
over a volume in momentum space,
corresponding   to    the    composite    wave-function,
exceeds a phenomenological cutoff-value of 0.30.
At low densities, in the limit
of many interactions, the set of equations for the composites
and nucleons yields the
required law of mass action.  A coalescence
scaling~\cite{nag84} for the composite
spectra is expected to emerge from the model based on the
equations, when
the processes of composite production do not vary substantially
with space and time.

The pions within the model are produced in two
steps~\cite{yar79}.  First a
$\Delta$ or $N^*$ resonance is produced in an $NN$ collision,
and then the resonance decays into a nucleon and pion.  The
cross section for the resonance production appears to
nearly exhaust the inelastic $NN$ cross
section in the energy range of interest~\cite{ver82}.
The
resonance occupation functions
satisfy transport equations that
follow from the nonequilibrium many-body theory in the adiabatic
limit~\cite{dan91},
\begin{eqnarray}
{\partial \over \partial t} \, \big( f_X\,{\cal A}_X\big) +
{\partial E_X
\over \partial {\bf p}_X} {\bf \cdot} {\partial \over
\partial {\bf r}} \, \big( f_X\,{\cal A}_X\big) - {\partial E_X
\over \partial {\bf r}} {\bf \cdot} {\partial \over
\partial
{\bf p}_X} \, \big(f_X\,{\cal A}_X\big) = {\cal K}^<_X (1 \mp
f_X) {\cal A}_X - {\cal K}^>_X f_X {\cal A}_X , \label{rboltz}
\end{eqnarray}
where ${\cal A}_X$ is the spectral function describing the spread
of the resonance in mass,
\begin{equation}
{\cal A}_X = {\Gamma_X \over (m - m_X)^2 + {1 \over 4}\Gamma^2_X},
\end{equation}
and $\Gamma_X$ is the resonance width which generally depends
on $m$.  The derivatives on the l.h.s. of Eq.~(\ref{rboltz}) are
taken at a constant $m - m_X$.   The absorption of pions is
described
with a~sequence of inverse processes to those in the
production. The resonance formation dominates the $\pi N$ cross
section at the energies of interest, with the cross section
for resonance $X$ of the form~\cite{pil79},
\begin{equation}
\label{piNcs}
\sigma_{\pi N\rightarrow X} = {\pi \over p^2}
{g_{X}\over
2} \Gamma_{X \rightarrow \pi N} {\cal A}_{X}
\end{equation}
with $p$ - c.m. momentum
and $g_X$ - resonance spin-degeneracy.  Microscopic
reversibility and the assumption of a weak dependence of
the interaction matrix-element squared on the resonance mass,
yield a relation between the cross sections for
resonance absorption and production~\cite{dan91,bao93},
\begin{equation}
\sigma_{XN\rightarrow  N'N"} = {1 \over 1 +
\delta_{ N'N"}} {2 \over g_X} {m p^2_{ N'N"} \over
p_{XN}} {\sigma_{{ N'N"}\rightarrow X {
N}} \over \int^{E-m_N}d \widetilde{m}
{\cal A}_{X}(\widetilde{m})\widetilde{p}_{XN}},
\end{equation}
where $E$ is total energy.
A variant~\cite{wol92} of the relation relies on the
assumption
of a rapid variation of the matrix element, for $m$ close
to $E - m_N$.
In the calculations preceeding~\cite{dan91}, a
detailed balance relation between the resonance
production and absorption cross-sections used was such as for
stable particles, leading to an excessive number of pions in
equilibrium.  In the pion-nucleus interactions the true pion
absorption cross section was underestimated.
Further violating reversibility in the
calculations, in (\ref{piNcs}) the momentum prefactor was
replaced by a constant, affecting the shape of pion
spectra.

The single-particle energies in Eqs.~(\ref{boltz})
and~(\ref{rboltz}) are parametrized
in terms of an optical potential, with a~local part handled
as a scalar potential,
and a weaker nonlocal part added on to the particle
energies in the system c.m.,
\begin{equation}
E_X  =  (p^2 + m^2)^2 + q_X\,\Phi + A_X\,U^{C}_1, \hspace{3em}
m =  m_0 + A_X\,U^C + t_{3X}\,U^T,
\end{equation}
where $q_X$ is particle charge, $\Phi$ -- Coulomb potential,
$A_X$ -- particle mass number, $m_0$ -- mass in free space,
and $t_{3X}$ -- isospin component.  The dependence of the
potentials $U$ on scalar densities is chosen such as in
nonrelativistic calculations,
\begin{equation}
\label{Us}
U^C  =  -a \, {\rho_{s} \over \rho^{0}} + b\left( {\rho_{s} \over
\rho^{0}}\right) ^{\nu} , \hspace{3em}
U^{C}_1 = -d \, \nabla^2 \left( {\rho_s \over \rho^0} \right)  ,
\hspace{3em}  U^T = c\, {\rho^{T}_{s} \over \rho^{0}} ,
\end{equation}
where
\begin{equation}
\rho_{s}  =  \sum_{X} A_X \rho_{sX} , \hspace{3em} \rho^{T}_{s} =
\sum_{X} t_{3X}\rho_{sX} ,
\end{equation}
with
\begin{equation}
\rho_{sX} = g_{X} \int {d{\bf p} \over (2\pi )^{3}}
{m_{X0}\over E_{X}} f_{X} , \hspace{1.5em}\mbox{and}\hspace{1.5em}
\rho_{sX} = g_{X} \int {d{\bf p} \over (2\pi )^{3}} \int {dm
\over 2\pi} {m_{X0}\over E_X} f_{X}\,
\end{equation}
for the stable particles and
resonances, respectively.  The neutron and proton density profiles for
initial
nuclei are determined by solving a differential
equation for baryon density that follows from the Thomas-Fermi
(TF)
equations in the nuclear frames,
\begin{equation}
E_p(r,p_{Fp}(r)) = \mu_p,\hspace{3em} E_n(r,p_{Fn}(r)) = \mu_n.
\end{equation}
The chemical potentials $\mu_p$ and $\mu_n$ are adjusted till
the values of total
proton and neutron numbers for nuclei are reproduced.  It
appears
convenient to solve differential equations set up for the
proton
and neutron numbers, Coulomb potential, and net energy in
parallel to the equation for baryon density.

The normal
density  is taken equal to $\rho_0 = 0.160$ fm$^{-3}$.  The
description of charge
densities
from electron scattering~\cite{jag74} is better with this
value than with $\rho_0 =
0.145$ fm$^{-3}$~\cite{dan92}.
As~to the other parameters in (\ref{Us}),
$a=357$ MeV, $b=304$~MeV, and
$\nu={7 \over 6}$ correspond to the soft equation of state ($K
\simeq 200$ MeV), and $a=123$~MeV, $b=70.1$~MeV, and
$\nu=2$ to the stiff equation ($K
\simeq 375$ MeV); $c = 92$ MeV,
$d=22.4$~MeV$\,$fm$^2$~\cite{len89,rin80}.  Sample densities for the second
parameter set are shown in~Fig.~\ref{profil}.  Unless otherwise
indicated other results in the paper are obtained using the
second set.  The effects of the momentum dependent
potentials
on the expansion have not been explored as of yet, due to
the involved numerical difficulties with energy
conservation and a specific
interest in the differences between mean energies and spectra
of different emitted fragments.

Occupation functions in the phase space for stable particles,
and in the phase space and mass for resonances, are
represented in the calculation with test particles.  The drift
terms on the l.h.s. of Eqs. (\ref{boltz}) and (\ref{rboltz}) are
integrated by requiring that the test-particle positions and
momenta satisfy the Hamilton's equations.  A lattice
hamiltonian method of Lenk and Pandharipande~\cite{len89} is
used with the scalar density at lattice sites evaluated
with a form factor
\begin{equation}
{\cal S}_X({\bf r},p) = {A_{X} \over
N_{test}(2\ell)^3} {m_{X0}\over E_X}  g(x)g(y)g(z),
\end{equation}
where $g(q)
= 1$ for $|q| < 0.5\ell$, $g(q) = 1.5- |q|/\ell$ for $0.5\ell
< |q| < 1.5\ell$, $g(q)=0$ for $|q| > 1.5\ell$.  Further,
$N_{test}$
is the number of test particles per particle, and $\ell$ is
lattice spacing.  The particular form of $g$ allows for a good
total energy conservation (to within 0.25 MeV/nucleon in
a 100 MeV/nucleon collision), and a good momentum conservation
for
asymmetric systems, with no considerable computational effort.
It should be mentioned that an initialization of the system
according to the densities from the TF equations is essential
for an "elastic" algorithm for the drift terms.  For other,
however sensible, initializations, isolated nuclei
develop
strong breathing oscillations that can deplete central density
to a~quarter of normal value.
 The Coulomb potential is found
by solving the Poisson equation using a novel relaxation
method~\cite{fel92}.  Integrals in the collision rates in Eqs.
(\ref{boltz}) and (\ref{rboltz}) are computed using the
Monte-Carlo method~\cite{dan91} (see also Ref.~\cite{lan94}).
Entropy is calculated by integrating~\cite{dan91} the~rate of
entropy variation due to collisions.

\section{HEAD-ON COLLISIONS OF HEAVY NUCLEI}
\label{Central}


Emphasis is going to be on symmetric reactions of heavy nuclei.
For symmetric systems the experimental identification of specific
collective effects
may be less ambiguous than for
asymmetric systems, due to the well defined center of mass
and the lack of shadowing at lowest impact parameters.

\subsection{Shock Fronts}

The dynamics of the central high-energy reactions 
can be broken down into several well identified
stages.  A 400 MeV/nucleon Au + Au system at $b = 0$
will be used 
to illustrate specific points.
Contour plots of baryon density at different times for that
system are shown in the left column of Fig.~\ref{contours}.
Initially in the energetic reactions the nuclear densities
just
interpenetrate, with target and projectile nucleons separated
in the momentum space.  As~the time progresses, in the overlap
region
$NN$ collisions begin to thermalize the matter, making the
momentum distribution 
centered at zero
momentum in the~c.m.s.  Thereafter, the features of the matter
at the center of a system stabilize to a degree.
As further the region with the excited matter grows in size,
in a heavy system
interface regions can be identified
in-between the excited and
normal matter
where such parameters as baryon
and entropy density change rapidly, see Figs.~\ref{contours}
and~\ref{profau}.  In the head-on collisions the interfaces
are perpendicular to the beam axis.  With
the normal matter diving into the region with excited matter
at a speed well in excess of the speed of sound in normal
matter
(for the first sound the speed is $c_s = \sqrt{K/9m_N}$,
i.e.~0.15$c$  and 0.21$c$ in the case of soft and stiff equation
of state, respectively),
at the high beam energies,
the interfaces are recognized as
shock fronts.  In~the hydrodynamic limit a~discontinuity in the
velocity in the initial state, such as in-between the
projectile and target nucleons at $b = 0$, generally breaks
into two shock fronts travelling in opposite
directions~\cite{lan59}.
Figure~\ref{rankhu} displays the nuclear-matter parameters
within the
hydrodynamic limit,
behind a shock front bringing the matter to a~halt, as
a~function of beam energy, obtained by
solving the Rankine-Hugoniot (RH) equation~\cite{lan59}
\begin{equation}
{(e_1 + P_1)^2 \over \rho_1^2} - {e_0^2 \over \rho_0^2} - P_1
\left( {e_1 + P_1 \over \rho_1^2} + {e_0 \over \rho_0^2} \right)
= 0. \label{ranf}
\end{equation}
The parameters with subscript 1 in (\ref{ranf}) refer to the
shocked matter and $e$ and $P$ are the energy
density (including mass) and pressure, respectively.
Figure~\ref{rankhu} displays further the maximum densities
from the $b=0$
Au + Au simulations and these densities appear to follow the
general trend predicted by (\ref{ranf}).
The excited matter in-between shock fronts in the
simulations is exposed to the vacuum
in the sideward directions and, respectively, begins to expand
into these directions.  The features of this
process at high energies and at $b=0$ may be understood
at a {\em qualitative} level
in terms
of the self-similar cylindrically-symmetric hydrodynamic
expansion of an ideal gas.

\subsection{Expansion}
\label{expcen}

In the self-similar expansion~\cite{zel66,bon78}, the velocity
is proportional
to a distance from the symmetry axis, $v = {\cal F} (t)\,r$,
which
roughly appears to hold in the direction of 90$^{\circ}$ in
the Au~+~Au collision at 400 MeV/nucleon, cf.
Fig.~\ref{profau}.
Under the simplifying assumptions of an~isentropic flow and
an ideal-gas equation of state, yielding
$P/\rho^\gamma$ = const with $\gamma$ - the heat ratio, the solution to
nonrelativistic hydrodynamic equations for the density becomes
\begin{equation}
\rho (r,t)= \rho (0,0) \, {R^2(0) \over R^2(t)} \,
\left(1 - {r^2 \over R^2(t)} \right) ^{1 / (\gamma - 1)}.
\label{rhot}
\end{equation}
The radius $R$ of the density distribution satisfies
\begin{equation}
\ddot{R}(t) = {2 c_s^2(0,0) \over \gamma R(0) }\, \left({R (0)
\over R (t)}\right)^{2\gamma - 1},
\label{R2dot}
\end{equation}
with $c_s^2 = \gamma T/m$ - the speed of sound squared.  The
coefficient ${\cal F}$ is related to $R$ with ${\cal F}(t) =
\dot{R}(t)/R(t)$. Equation (\ref{R2dot})
may be partially integrated to yield, for $\dot{R}(0) = 0$,
\begin{equation}
\dot{R}(t) = c_s(0,0) \left[{2 \over \gamma (\gamma - 1)} \left(1 -
\left({R(0)\over R(t)} \right)^{2\gamma-1}\right)\right]^{1/2}.
\label{R1dot}
\end{equation}
Following Eqs. (\ref{R2dot}) and (\ref{R1dot}), the
coefficient ${\cal F}$ is expected to rise, first linearly, for
the time $t \sim R/c_s$ needed for a signal to propagate inward through the
radius, and it is expected to fall off later as the inverse of
time, reflecting the  quadratic initially and linear asymptotically
 dependence of $R$ on time.  Indeed, in~the high-energy simulations,
the velocity slope $dv/ dr$ in
the transverse direction, cf.~Fig.~\ref{profau},
behaves approximately as described.  Within the nonrelativistic
ideal-gas approximation, the speed of sound in the central
region of the collision at 400 MeV/nucleon is assessed, with
$\gamma = 5/3$, at $c_s \sim 0.4 c$.
The radius of the gold nucleus is $\sim
7$ fm, and thus the slope might be expected to rise for
a~time~$\sim 18$~fm/c.  The~maximum in
the simulations is reached at  $t \sim 27$ fm/c which appears
consistent with the simple estimate as initially some time is spent
equilibrating and compressing matter, cf.~Fig.~\ref{densau}.
The~value
at maximum $dv/dr \approx 0.040$ c/fm is not far from the one
obtained by integrating~(\ref{R1dot}), ${\cal F} \approx
0.6\,
c_s/R \sim 0.036$ c/fm.   At the given instant, according to
(\ref{R1dot}),  the~transverse matter radius
should be about 50\% in excess of
the original which approximately is the
case in the simulation, cf.~Figs.~\ref{contours} and
\ref{profau}.  After the
time $R/c_s$, the density in the central region of the reaction
should decrease as $t^{-2}$, following Eqs. (\ref{rhot}) and
(\ref{R1dot}).

While the expansion develops in transverse directions,
the normal matter continues to enter the excited region along
the beam axis for a time $\sim (2R/v_0)(\rho_1 -
\rho_0)/\rho_1 \sim 16$ fm/c in the Au + Au collision.  Here
$v_0 \approx 0.42\,c$ is initial
velocity in the center of mass and $\rho_1$ is  the density in
the shocked
region.  After the shock fronts reach the vacuum, an expansion
{\em along} the beam axis sets in that requires a~time
$\sim (2R/c_s)(\rho_0/\rho_1) \sim 18$ fm/c for a~full
development.  By~then, however,
the matter is already quite decompressed with
particles
having moved out in the transverse directions.  At the later
stages, with the expansion becoming three-dimensional, the
density
in the central region can be expected to drop~as~$t^{-3}$.

While the discussed and few other dynamic features
of $b=0$ high-energy reactions
appear to be consistent with a hydrodynamic behavior of the
matter, there
are also definite differences which cannot be classified as
effects of the
mean-field potential or relativity.
Thus, the baryon density at the center of the system,
Figs.~\ref{profau}, \ref{rankhu}, and \ref{densau},
falls short of the value of $2.5 \, \rho_0$ expected at 400
MeV/nucleon on the basis of the RH equation.  The~entropy per
baryon never quite reaches values expected for an
equilibrated
system at the center.  Typical combinations of density and
mean nucleon kinetic energy such as $\rho \sim 2.1 \, \rho_0$ and
$E_{kin} \sim 91 \, \mbox{MeV}$, should result in $S_B/A
\sim 2.8$, compare Fig.~\ref{profau}.  For reference,
the~nonrelativistic formula for the entropy, with statistics
ignored, is
\begin{equation}
{S_B \over A} = 2.5 - \log \left( {\rho \over 4} \left( {3 \pi
\hbar^2 \over m_N E_{kin}} \right) ^{3/2} \right) .
\label{sba}
\end{equation}
The shock fronts do not fully
develop~\cite{dan84} and/or the system never gets completely
thermalized at the center (see also~\cite{cas90}).
In a system
expanding hydrodynamically into
vacuum, local temperatures would have dropped to zero  and  all
kinetic energy would have turned into the collective energy of
expansion~\cite{zel66,bon78}.  For that,
collisions would need to continue down to appropriately low
densities.  The mean density in the vicinity
of last collisions of particles emitted from head-on Au +
Au reactions,
is shown as a function of the beam energy in~Fig.~\ref{frezen}.
At~400~MeV/nucleon the freeze-out density is equal to about
$0.2\rho_0$.
The~bottom panel in Fig.~\ref{densau} shows the evolution of
the participant collective transverse energy per nucleon in the
400~MeV/nucleon reaction, which is
calculated from the perpendicular components of a~local
hydrodynamic velocity  ${\bf v}$
in the vicinity of last collisions,
\begin{equation}
E^{\perp }_{coll}/A = \sum_X \int d{\bf r} d{\bf p} f_X
m_X (v^{\perp })^2 \left/
\sum_X A_X\int d{\bf r}d{\bf p} f_X  \right. .
\label{Ecol}
\end{equation}
The collective transverse
energy per nucleon freezes out in Fig.~\ref{densau}
at a value of $~42$ MeV.  For~nucleons
this constitutes about 50\% of the kinetic energy in
transverse degrees of freedom at the freeze-out.  (A bit
larger fraction is found if relativistic effects are taken into
account.)  The nuclear systems are, indeed, too
small for the entire energy to turn into the collective energy.
Simple estimates of mean freeze-out density based on
\begin{eqnarray*}
\langle \rho \rangle \simeq \int dt\;{\rho \over \tau} \exp
\left(\int^t{dt'\over \tau } \right)
\end{eqnarray*}
where $\tau = (\rho \sigma_{NN} v^*)^{-1}$ and $v^*$ is
relative
thermal velocity, that utilize a nuclear density $\rho(t)
\propto t^{-\nu}$
where $\nu
= (2-3)$, and rely  on the assumptions of either adiabatic
or isothermal expansion yield $\langle \rho \rangle /\rho_1
\propto (\sigma_{NN} R \rho_1)^{-\kappa}$, where
$\rho_1$ is the density at the start of expansion
and
$\kappa$ ranges from~2/3 to 2.  Following these estimates, the energy
dependence of the freeze-out density seen in~Fig.~\ref{frezen},
above 100 MeV/nucleon,
may be attributed to the variation of elementary
cross sections of particles interacting locally with
reduced relative velocities.  If~cross sections are
made energy-independent in simulations, indeed little
change in $\langle \rho \rangle$ is observed towards  high
energies.

One of the effects of transverse expansion on observables
from very central collisions in the simulations, is the peaking
of particle
polar-angle distributions at $90^{\circ}$ in the c.m.s.
This peaking is more pronounced in the
distributions of clusters which are
more affected by the hydrodynamic motion than are
nucleons~\cite{dan79}, see e.g. Fig.~\ref{polau}.  With the variation of
beam energy, the peaking
is observed to maximize at about 150 MeV/nucleon.
Below~40~MeV/nucleon the emission turns isotropic.  The~shock
width at low energies becomes comparable to the nuclear
size~\cite{dan84} and thus a~system equilibrates as a~whole.
For polar-angle distributions
from the simulations of 250~MeV/nucleon reactions see
Ref.~\cite{dan92}.  At~very high energies (and/or for light
systems like~Ca~+~Ca) the effects of expansion in the
simulations
compete with the transparency effects in the corona.
The proton distributions become forward-backward peaked
in~the~c.m.s., at beam energies above 1.5 GeV/nucleon,
and the light-cluster distributions above 1.9~GeV/nucleon in
the Au~+~Au reactions.  As~seen in Fig.~\ref{polau}, the
pion polar-angle distribution peaks at $\theta = 0$ already
at~1~GeV/nucleon.  For~the purpose of
demonstrating that, indeed, the transparency plays a role, the
figure shows, in~addition to the distribution of all pions,
the distribution of only those pions which were emitted within
the first 14 fm/c.
The~latter pions constitute about 1/3 of all, and the
anisotropy of their polar-angle distribution is nearly 3 times
larger than
the anisotropy for~all.  (Possible effects of geometry in
a~surface
emission in shaping the pion angular distribution may be
eliminated,
as the angular distributions at lower energies are peaked at
90$^{\circ}$.)
Despite the transparency
effects
in the corona, the maximum densities at the center of the Au +
Au system are not farther away from the RH expectations at the
highest of the studied energies than at lower,\footnote{In a
comparison in \protect\cite{dan92} over a more narrow range of
energies, the pions and baryon resonances were omitted in the
RH prediction and some nonrelativistic approximation was made.
Besides, the
predictions there were for another equation of state.}
cf.~Fig.~\ref{rankhu}.
The~peaking of particle distributions
at~90$^{\circ}$ at $b=0$, including that for pions at lower
beam energies, eventually disappears with an
increase of impact parameter~\cite{dan92}.
Experimentally, the~most central collisions can be selected
by using the particle transverse energies or symmetry around
the beam axis~\cite{jeo94}.  In~order to avoid
a~bias in the angular distributions, the~analysed particles
should be
{\em excluded} from the trigger condition
(compare~\cite{dan85}).  By~forcing the condition
to a limit the extraction of information from most central
collisions
could be reduced to the issue of collecting an adequate event
statistics.

Figure \ref{eneau} shows
the mean transverse and
longitudinal kinetic-energy components~\cite{dan92},
\begin{equation}
E_{kin} = E - m = \sqrt{m^{2} + p^{2}} - m =
{(p^{\perp})^{2}\over
m + E} + {(p^{\parallel})^{2}\over m + E} = E^{\perp } +
E^{\parallel},
\label{ekin}
\end{equation}
of particles emitted from the $b=0$ Au + Au collisions, as a
function of the bombarding energy.  The scale for the transverse
component is shifted downward by log 2 relative to
the scale for the longitudinal component, in order to allow for
a direct visual comparison
of the components {\em per degree of freedom} at the same
bombarding energy.
It is seen that, per degree of freedom, the transverse baryon
energies are higher than longitudinal in a~vast range of
bombarding energies.  This is a reflection of higher, per
degree of freedom, collective transverse energy than
longitudinal, at~freeze-out~\cite{dan92}.  The components of
the collective energy per nucleon, calculated as in
Eq.~(\ref{Ecol}), are shown in Fig.~\ref{colau}.
If particle momentum distributions are approximated with
anisotropic gaussians and nonrelativistic kinematics is
used, then the particle polar-angle distributions become
\begin{equation}
{4 \pi \over N} \, {dN \over d\Omega} = {(1 + \chi)^{1/2}
\over (1 + \chi \cos^2 \theta)^{3/2} },
\label{dNdOm}
\end{equation}
where $\chi = (\langle E^{\perp } \rangle /2 -
\langle E^{\parallel} \rangle )/\langle E^{\parallel} \rangle$.
Result  (\ref{dNdOm}) for tritons
emitted from a 1~GeV/nucleon Au~+~Au collision is shown by a
solid line in
Fig.~\ref{polau}.  In general, it is found that
Eq.~(\ref{dNdOm})
describes the distributions of emitted particles from
high-energy collisions
in a~semi-quantitative way, even the pion distributions.  The
90$^{\circ}$-to-0$^{\circ}$ ratios,
\begin{eqnarray*}
R_{\theta} = \left({dN \over d\Omega} \right)_{90^{\circ}} \left/
\left({dN \over d\Omega} \right)_{0^{\circ}} \right. ,
\end{eqnarray*}
are quite well described within the approximation, $R_{\theta} \approx
(1 + \chi)^{3/2} = (\langle E^{\perp
} \rangle /2 \langle E^{\parallel}\rangle )^{3/2}$.
Within the gaussian
approximation, the mean particle energy at 90$^{\circ}$ is
equal~to~\cite{dan92}  $1.5 \times \langle E^{\perp
} \rangle$.  The~mean energies
at 90$^{\circ}$ in simulations are slightly
lower.

\section{SEMICENTRAL COLLISIONS}
\label{Semi}

The attention is now turned to the dynamics of semicentral
collisions, with emphasis put on the similarities and
differences compared to the dynamics at $b=0$.

\subsection{Hydrodynamic Discontinuities}

The initial discontinuity between the projectile and target
velocities breaks at a finite~$b$ in the
ideal-fluid limit into  two shock fronts
propagating into the projectile and target, {\em and}~a~weak
tangential discontinuity in-between~\cite{dan79,lan59}.
Unlike across a shock front, the~normal velocity component,
pressure and, in the case at hand, entropy and density are
continuous across the tangential discontinuity, and
the tangential velocity component is not.  The~discontinuities
start out in collisions all at the same angle
$\alpha$ relative to the beam axis,
cf.~Figs.~\ref{sketch}~and~\ref{contours},
\begin{equation}
\cos \alpha \approx {b \over 2 R} \, {1 \over \sqrt{ 1
+(E_{beam}/2m_N) (1 + b^2/4 R^2)}} ,
\label{alpha}
\end{equation}
where the beam energy is in MeV/nucleon, and the Lorentz
contraction in the c.m.s. is taken into account ($\,v_0 = (1 +
2m_N/E_{beam})^{-1/2}$, $\gamma_0 = \sqrt{1 +
E_{beam}/2m_N}\,$).  Developing inclined shocks can be
seen in the $t = 10$ fm/c panels of Fig.~\ref{contours} for a
finite $b$.  Signs of a~tangential discontinuity for a
finite $b$ are visible in the densities at later times in the
figure.  Figure~\ref{tang} shows finally the variation of
different  quantities along a normal to discontinuities in
a~collision at $b = 7$~fm.  Note that the center of the weak
discontinuity is at $r = 0$ where~$v_t$ vanishes, while the
center of a shock is at $r \sim 3.1$ fm
where $v_n$ reaches 1/2 of the value for the intact matter.

In the ideal-fluid limit, the tangential component of a
velocity would be unaltered by a~shock, and thus the matter
behind the shock in collisions would be
expected to move at a~speed $v_1 = v_0 \cos \alpha$ and
angle $\alpha$ relative to the beam axis.  The change in the
thermodynamic parameters should be then such as
for a shock perpendicular to the beam axis, at~$b=0$ and a beam
energy (called further an equivalent energy)
\begin{equation}
E_{beam}' = E_{beam} \sin^2 \alpha ,
\label{ealpha}
\end{equation}
where relativity has been  thoroughly accounted for.  The
ideal-fluid dynamics
predicts then that the maximum density reached early on in
collisions in the c.m. at a finite  $b$, is related
to the density at $b = 0$ at a lower beam energy, by
\begin{eqnarray}
\rho_1 (b, E_{beam} ) & = & \gamma_1 \rho_1
(0,E_{beam}') \nonumber \\[.1in] & = &
\left(1 - {\beta \epsilon \over (1 + \epsilon) (1 + \epsilon (1
+ \beta))} \right)^{-1/2} \rho_1 \left(0,E_{beam} {(1+
\epsilon) (1 - \beta) \over 1 + \epsilon (1 + \beta)} \right)
\nonumber \\[.1in] & \approx & \rho_1 (0,E_{beam}(1 - \beta)) = \rho_1
(0,E_{beam}(1 - b^2/4 R^2)),
\label{ranr}
\end{eqnarray}
where $\epsilon = E_{beam}/2m_N$ and $\beta = b^2/4R^2$.   The
approximation following the equalities in~(\ref{ranr}) is
nonrelativistic ($\epsilon \ll 1$).  The c.m. maximum density
and corresponding rest-frame parameters like entropy behind the
shock fronts are
expected to change slowly according
to~(\ref{ealpha})~and~(\ref{alpha}), only quadratically,  with
the rise of $b$.  Indeed, a weak variation behind the shock
fronts is found in the simulations.
The validity of the relativistic relation~(\ref{ranr}) is tested
in the top panel of Fig.~\ref{rankhu} that displays, in
particular, the maximum density from 1~GeV/nucleon Au~+~Au
collisions at different $b$, divided by $\gamma_1$, vs
$E_{beam}'$ from~(\ref{ranr}).  The relation appears to work up
to $b \sim 11$ fm at this and other beam energies.
Further predictions based on the ideal-fluid dynamics need not
work as well.

In~the simulations all the discontinuities
have widths that are quite comparable to the size of a
system, see Fig.~\ref{tang}.  The tails of the weak and strong
discontinuities
overlap and in consequence the velocity in-between does not, in fact,
reach
values such as if the discontinuities were separated.
For example, at low $b$, along the axis
passing through the center, the magnitude of the transverse
velocity component in the reaction plane reaches at most 40\%
of $(v_0 \sin \alpha  \cos \alpha)$ expected in-between
separated discontinuities.  As a function of time,
the width of any shock front (of the order of a~mean-free
path for a strong shock)
remains constant~\cite{dan84,lan59}, provided that the
conditions on the two
sides of the front remain the same.  The width of a
tangential
discontinuity, on the other hand, increases with time.
That this must be the case can be seen directly by adopting
the hydrodynamic Navier-Stokes equations~\cite{sre77}\ to a
region of
the discontinuity, where they yield a diffusion equation for
the tangential
component of velocity (or for vorticity $\omega =
{\bf \nabla} \times {\bf v}$). The diffusion
coefficient  is equal to
 $\eta/(m_N \rho)$, where
$\eta$ is the shear viscosity coefficient (proportional, in
general, to the mean-free path).  Solving this diffusion
equation yields
the tangential velocity component as a function of time $t$ and
distance $r$ in a~direction perpendicular to the
discontinuity, in the form
\begin{equation}
v_t(r,t) = v_t^0 \, \mbox{erf}\left({r \over 2} \sqrt{{m_N \rho
\over \eta t}} \right) ,
\label{tan}
\end{equation}
where $v_t^0$ is
an asymptotic value of the velocity component and erf is error
function.
According to Eq.~(\ref{tan}) the half-width of a weak
discontinuity should rise with time as $\delta = 2 \sqrt{\eta t
/ m_N \rho}$.
For the situation in~Fig.~\ref{tang}, with $\eta \sim 50$
MeV/fm$^2$c~\cite{dan84}, this yields $\delta \approx 3.2$ fm
which agrees well with what can be deduced
from the simulation in the figure.
Generally, during the time span lasting from the initial
equilibration in simulations
till the development of the expansion at intermediate and low
impact parameters, the tangential-velocity profile at the
center, is reasonably well described by Eq.~(\ref{tan}),
despite of the proximity
of shocks.
The~root in (\ref{tan}) alleviates uncertainties associated
with  the choice of density and a reference time.  While
$\delta$
increases as $t^{1/2}$, the separation between shocks increases
as $t$, and thus, if nuclear systems were
large enough and/or viscosity low, the shocks in collisions could
separate
well from the weak discontinuity, before reaching the vacuum.
For large Reynolds
numbers ${\cal R} = m_N \rho v  R/\eta$
the weak
discontinuity at the center could
break into a turbulence~\cite{lan59}.   In~the
system studied ${\cal R} \sim 9$,
while typical critical values for turbulence are
larger~\cite{lan59}.  No~sign of   any turbulence
is  seen in the simulations.

In the course of a reaction, the section of the weak
discontinuity
turns from one given by a~straight line to one reminding a
hyperbolic tangent, and on the average the discontinuity
rotates, decreasing an angle relative to the
beam axis.  That is because the
matter moving forward on one side of a reaction plane, away
from the center of the system, emerges facing
little or no target matter moving the other way.  Symmetrically,
the target matter on the other side emerges facing  little or
no projectile matter.
The~shock fronts subsonic relative to the matter
behind rotate as well.  The fronts are generally strong in the
vicinity of an axis passing through the center of a system
in the direction of the beam; there from the~shock in
projectile
matter continues as strong towards the target side of
the reaction plane, and the shock in target matter as strong
towards the projectile side of the reaction plane.  With the
excited region within the geometric overlap of projectile
and target fed
by incoming matter, the density is maintained behind the
shocks during the shock propagation.

The~shock
in projectile matter is weakest in the vicinity of its edge on
the projectile side of the reaction plane.  There the front
curves, making a low angle relative to the beam axis, the
change
in the velocity component normal to the front is low, and
correspondingly the equivalent energy for the front is low,
compare~(\ref{ealpha}).  Symmetrically, the front moving into
target
matter is weakest in the vicinity of its edge on the target
side of the reaction plane.
The
distinction between the strong and weak portions of the
fronts is generally pronounced at high beam energies but less
at low. The weak shocks propagate with only
slightly supersonic velocities relative to the intact matter.
These shocks, in what becomes spectator matter at high
energies,
might remind those away from supersonic bodies moving
through a~medium.  The~matter little disturbed by the weak
shocks can move
out as an~entity  largely in its original direction
in the c.m., see the $t = 35$ fm/c panel in Fig.~\ref{contours}
for $b = 7$ fm.   (The spectator matter separates out also in
the
reaction at $b = 3$ fm at a later time than shown in the
figure; the shapes for the matter are far from spherical.)

 If~the~hydrodynamic discontinuities in the
region
of geometric overlap between projectile and target, were
more separated in the simulations, the matter would have
moved more to the sides within the reaction plane behind
the shocks,
strengthening the shock edges.  The~spectator matter would not
have been then as distinct; see
the results~in~\cite{ams77}. Weakness of
the sideward flow in collisions~\cite{dan88a}, as compared to
the
ideal-fluid predictions, and the emergence of spectator matter
are thus very directly related.

\subsection{Expansion}
\label{expb}

{}From the shock formation on, an expansion proceeds
to develop in-between the fronts.  At~low~$b$ the
expansion in the directions perpendicular to the reaction plane
has
features similar to the expansion in the directions
perpendicular to the beam axis at $b = 0$.  The~variation of baryon
density, entropy and radial velocity at different time
instants, with the distance from the
center of the system along a line defined by $\theta =
90^{\circ}$
and $\phi = 90^{\circ}$, where $\phi$ is the azimuthal angle
relative to the reaction plane, is quite close to
the variation given by the dashed lines in Fig.~\ref{profau}.
The top panel in
Fig.~\ref{enrec} shows the dependence on the impact parameter
of the mean particle kinetic energy components
out of the reaction plane
$E^y =  (p^y)^2 / (m + E)$,
where $p^y$ is
the momentum component out of the reaction plane, at c.m.
rapidity $y = 0$.  It~is seen that the mean energy
components depend
roughly quadratically on $b$ for low values~of~$b$ (i.e. in
such a manner as the maximum
density in the collisions).   Center panel in the figure shows
the energies $E^y$ averaged over whole momentum space.  These
averages are equal to half of $\langle E^{\perp} \rangle$ at $b
= 0$, and they decrease faster with an~increase of $b$ than
the averages at $y=0$, because of larger spectator
contributions.
Either type of average energies may be determined
experimentally,
correcting for fluctuations of the reaction
plane~\cite{dan87,dem89,dan85}.

With regard to the directions in the vicinity of the reaction
plane,
the initial velocities tangential to the weak discontinuity
behind shock fronts (although low compared to those in the
ideal-fluid
limit) add on to the velocities developed in expansion, and
affect the sideward-flow
observables.  The deflection of particles moving forward
and backward in the c.m. is commonly quantified in terms of the
mean momentum in the reaction plane as a function of the
rapidity~\cite{dan85}.  Figure
\ref{pxa} shows the mean momentum divided by the mass, $\langle
p^x (y) \rangle /m$,  for particles emitted from Au + Au
reactions at 1 GeV/nucleon at two impact parameters.  For~the
purpose of minimizing the effects of
spectator particles at high energy, the slope of the mean
momentum at midrapidity is further used in
analyses~\cite{gus88}.  The slope
\begin{equation}
F =  {d \langle p^x/m \rangle \over dy}
\label{four}
\end{equation}
is shown in Fig.~\ref{slope}
for 400 MeV/nucleon Au + Au collisions, as a~function of the
impact parameter, both for the calculations and data.
This slope generally increases with particle mass.
If the
emission
pattern at low $b$ and low c.m. rapidities differed from that
at $b = 0$ {\em only} by a rotation of the plane of preferred
emission,
from the angle of 90$^{\circ}$ relative to the beam axis, to an
effective angle $\alpha$, then the
slope $F$ at midrapidity should be
approximately~\cite{dan87,sch87}
\begin{equation}
F \approx { \chi \sin \alpha \cos \alpha \over 1 + \chi
\cos ^2 \alpha } ,
\label{Fappro}
\end{equation}
with $\chi$ obtained from the energy components at $b=0$, cf.
(\ref{dNdOm}).
In obtaining (\ref{Fappro}) nonrelativistic kinematics was
used and
momentum distribution in the vicinity of midrapidity was
approximated in a gaussian form.
The maximum
value of $F$ from (\ref{Fappro}) as a function of $\alpha$ equals to
$\chi/(2\sqrt{1 + \chi})$.  For protons from
400 MeV/nucleon collisions this is 0.22 (using energy values
from Fig.~\ref{colau}).  However, the proton $F$ in
Fig.~\ref{slope}
reaches $\sim$0.47 at intermediate impact parameters,
indicating the importance of the weak discontinuity.
Note that the spectator contributions would tend to lower $F$.

On~taking spectator contributions into consideration, and on
analyzing the behavior of the r.h.s. of Eq. (\ref{Fappro}) with
$\chi$, one finds that $F$ provides a lower limit to the
angle of inclination of the discontinuities within participant
matter relative to the beam axis,
\begin{eqnarray*}
\alpha > \tan^{-1} F.
\end{eqnarray*}
The calculations in Fig.~\ref{slope}, utilizing
parameters for the
stiff equation of state and free NN cross-sections, actually
somewhat
overestimate data.  Thorough discussions of the dependence
of observables characterizing the sideward deflection on
the microscopic input parameters in simulations can be
found in Refs.~\cite{gal87}.  It follows from Fig.~\ref{slope}
that the highest multiplicity interval within the
experiment~\cite{gus88} corresponds to $b \,
\raisebox{-.5ex}{$\stackrel{<}{\scriptstyle\sim}$ } 2.5$ fm.

Either data or calculations may be analyzed in terms of
a kinetic-energy tensor~\cite{gyu82}.
This tensor can be written in~a~relativistic form as
\begin{equation}
S^{ij} = \sum_{\nu} s_{\nu}^{ij} = \sum_{\nu} {p_{\nu}^i
p_{\nu}^j \over m_{\nu} + E_{\nu} }
\label{sij}
\end{equation}
where the sum is over particles from a collision.
Experimentally, the
mean elements of the
particle tensors
$\langle
s^{ij} \rangle$, associated with the reaction plane, may be
determined directly as outlined in~\cite{dan87},
irrespective of the
fluctuations of an estimated reaction-plane about the
true plane.\footnote{Following considerations such as
in Ref.~\protect\cite{tsa91}, optimal weights in a vector
\protect\cite{dan85} for the estimation of the reaction-plane
direction,
${\bf Q} = \sum_{\nu} w_{\nu} {\bf p}_{\nu}^{\perp}$, are
$w_{\nu} \approx \langle p^x \rangle / \langle (p^{\perp})^2
\rangle$, where averages refer
to specific rapidities and particle types.  A more general
equation for the estimation of the azimuthal angle~$\Phi$ of
the reaction plane follows from the requirement $0 = {\partial
\over \partial \Phi} \sum_{\nu} ((p^{\perp}_{\nu}
\cos (\phi_{\nu} - \Phi) - \langle p^x \rangle )^2/ \langle
(p^x - \langle p^x \rangle )^2 \rangle + (p_{\nu}^{\perp} \sin
(\phi_{\nu} - \Phi))^2/\langle (p^y)^2 \rangle )$.  The
averages within the equations may be determined
self-consistently
in data analysis.} The major axis of the mean particle-tensor
points in a direction within the reaction plane, at an angle
$\theta_f$ relative to the beam axis.  This direction is
intermediate between the direction of discontinuities
within the participant matter and
the direction of motion of the spectator matter.
Eigenvalues
associated with the axis pointing out of the reaction plane are
identical with the mean energies $\langle E^y \rangle$.  The
minor axis
within the reaction plane, denoted with symbol $x'$, is
generally at a considerable angle
$\raisebox{-.5ex}{$\stackrel{>}{\scriptstyle\sim}$ } 50^{\circ}$
relative to the strong and weak
discontinuities within the participant matter.  As $b
\rightarrow 0$ this axis becomes identical with the
beam axis.
Eigenvalue energies $\langle
E^{x'} \rangle$ are generally less than~$\langle E^y \rangle$,
see~Fig.~\ref{enrec}.  The ratios
$\langle E^y \rangle / \langle E^{x'} \rangle$ increase with
particle mass.
These ratios maximize at $b
= 0$
as a~function of the impact parameter,
where they coincide with $\langle E^{\perp} \rangle /2
\langle E^{\parallel} \rangle $.
Compared e.g.~to~the variation of the flow angle $\theta_f$
(see also Ref.~\cite{sto86}), the variation of
the discussed energy ratios with $b$ is, though, slow,
cf.~Fig.~\ref{enrec} and further Fig.~\ref{rla}.

Kinetic energy tensor, in its
nonrelativistic form, was determined in the Au + Au reactions
on an event-by-event basis
in Ref.~\cite{gut90}.  The mean ratio $R_{\lambda}$ was
calculated of matrix elements
associated with the direction out of the estimated reaction
plane and the direction within the plane, perpendicular to
main axis, as a function of participant proton
multiplicity.  Fluctuations of the estimated reaction plane
about the true reaction plane generally reduce the possible
differences
between matrix elements in question~\cite{dan85}.  (If not for
fluctuations, $R_{\lambda}$ would have been identical with the
energy ratio $\langle E^y \rangle / \langle E^{x'} \rangle$.)
For~the~Au~+~Au reaction at 400 MeV/nucleon, the average
azimuthal angle between the two planes has been assessed within
the
experiment, and given in Ref.~\cite{gut90} as function of
proton
multiplicity.  This makes possible a practical comparison of
present calculations with these data.  Ratio~of the
eigenvalues
from the tensor with matrix elements modified on account of
the fluctuations,\footnote{The difference $(\langle s^{xx}
\rangle - \langle
s^{yy} \rangle)$ is reduced by a factor $\langle \cos
2\delta\Phi \rangle$
and the element $\langle s^{xz} \rangle$ is reduced by $\langle
\cos \delta\Phi \rangle$,
where $\delta\Phi$ is azimuthal angle between the estimated and
the true reaction plane~\cite{dan85,dem89}.} is~compared to the
data~\cite{gut90} in Fig.~\ref{rla}.  As in the case of
Fig.~\ref{pxa} a general agreement is found, although,
quantitatively the calculations overestimate somewhat the data.
The~lowering of the ratio $R_{\lambda}$ at high multiplicities
stems from an increase in fluctuations.
With~regard to the flow angle $\theta_f$, the position of a
maximum in the
experimental event distribution in the
angle~\cite{gut89a}, for the highest multiplicity bin, appears to be
in a rough accord with
the conclusion on the impact parameter range,
reached by examining $F$.

Physical reason behind larger energy values out of the
reaction plane in simulations, than in a direction within the
reaction plane, such as in the case of
Fig.~\ref{rla}, is the same as behind the different energies at
$b = 0$.  The expansion of largely equilibrated matter in
the~direction of motion of shock fronts is delayed,
compared to the direction out of the reaction plane, by the
time necessary for the shock fronts to traverse the normal
matter.
As~matter is already being decompressed by the expansion
developing in directions parallel to the shock-front surface,
the expansion in the direction of shock motion does
not acquire same strength.
The fact that data~\cite{gut90} (also~\cite{dan87,dem89}) show
different
energy values associated with different directions demonstrates
that, indeed, the equilibration occurs within
a part of the system separated out from the cold matter.  With
motion being generally supersonic at a high energy, the
interfaces
between the hot and cold matter {\em must have} the nature of
shock
fronts.  The~sideward flow stronger than anticipated on the
basis of a rotation of the $b = 0$ pattern, indicates the
presence
of a weak discontinuity (for the $b = 0$ pattern an
extrapolated energy ratio from finite-$b$ data such as in
Fig.~\ref{rla} could be used).

For other beam energies than 400 MeV/nucleon, azimuthal
deviations
between the estimated and true reaction planes were not
given~in~\cite{gut90}, and, correspondingly, the
measured and calculated energy ratios of interest in the Au +
Au reactions cannot
be directly compared.  Comparisons to the theory would have
been eased
if energies associated with the directions relative to the
{\em true} reaction plane were established, as was discussed.
Unless one deals with some artifact of a strong
variation of the fluctuations with beam energy, a~considerable
discrepancy between data and calculations, with
regard to the energy ratios, may be arising at low beam
energies. Specifically,
the
experimental squeeze-out ratio~$R_{\lambda}$~\cite{gut90}
maximizes
at 250~MeV/nucleon in central reactions and
rapidly decreases from there on with a lowering
of the beam energy.  On~the~other hand the ratio of calculated
energies  $\langle E^y
\rangle / \langle E^{x'} \rangle$ (which may be obtained
directly from~Fig.~\ref{eneau}), maximizes
at 150 MeV/nucleon and continues as significantly larger than~1
down to 50 MeV/nucleon.

\section{PION PRODUCTION}
\label{Pion}

It is generally believed that, in comparison to baryons, pions
are rather insensitive to the collective motion, due to their
low
mass and corresponding high thermal speeds.  First, briefly,
the possible
effects of a collective expansion on energies and spectra of
particles with different mass are investigated by
considering a situation of the instantaneous freeze-out in the
c.m.\footnote{More elaborate freeze-out conditions are e.g.
discussed in \protect\cite{sch93}.}
of a locally equilibrated system at a constant temperature.

\subsection{Collective Expansion and Mass}

With inclusion of particle statistics, phase-space
distributions in
a locally equilibrated
system are of the form
\begin{equation}
f_X = {1 \over \exp \left[(u^{\sigma }p_{\sigma } - \mu_X)/T
\right] \mp 1} = \sum_{n = 1}^{\infty} (\pm 1)^{n+1}
 \exp \left[ n( \mu_X - u^{\sigma }p_{\sigma })/T
\right]
,
\label{feqs}
\end{equation}
where the upper signs refer to Bose and the lower to
Fermi statistics, respectively,
$u^{\sigma } = (\gamma ,\gamma {\bf v})$, ${\bf v}$ is
velocity, and
$\mu_X$ is chemical potential.
The limit of a classical statistics corresponds to a~large
$(m_X - \mu_X)/T$.  In that case the first term in
(\ref{feqs}) dominates,
\begin{equation}
f_X \simeq \exp \left[(\mu_X - u^{\sigma }p_{\sigma })/T
\right] .
\label{feq}
\end{equation}
The corrections on account of statistics in (\ref{feqs})
correspond to an addition (for bosons) or subtraction (for
fermions) of classical components corresponding to lower
temperatures.  Of~a~primary concern below will be the classical
limit.

Nonrelativistically, a mean particle kinetic energy for a
locally equilibrated system may be represented,
following~(\ref{feq}), as consisting of thermal and collective components
in the form
\begin{equation}
\langle E^{kin}_X \rangle = \langle E^{kin}_X \rangle _{{\bf v}
= 0} + \langle E^{coll}_X \rangle = 3T/2 + m_X
\langle v^2 \rangle / 2 .
\label{avenre}
\end{equation}
The collective energy component rises proportionally to
particle mass, provided that $\langle v^2 \rangle$ is the
same for different particle species.
A relativistic extension of this result may be obtained by
expanding the relativistic mean particle energy in powers of a
collective velocity ${\bf v}$,
\begin{equation}
\langle E^{kin}_X \rangle =
\langle E^{kin}_X
\rangle _{{\bf v} = 0} + \langle E^{coll}_X \rangle
\approx
\langle E^{kin}_X \rangle _{{\bf v}
= 0} + (\langle E_X \rangle _{{\bf v} = 0} + 2 T) \,
\langle v^2 \rangle /2 + \cdots
\label{avere}
\end{equation}
The mean kinetic energy
of
particle
with mass $m$
in a {\em globally} thermalized system,
shown for reference in Fig.~\ref{ET}, is
$\langle E^{kin} \rangle _{{\bf v} =0}  = 3T + m (K_1 - K_2)/K_2$,
where $K_\nu$ is a modified Bessel function of the argument
$m/T$.
Relativistically, the thermal agitation increases the
coefficient in the collective energy component, which is a mass
in (\ref{avenre}), by the kinetic thermal energy plus twice the
temperature (i.e. at least by $7\, T/2$).  That is a large increase
for pions, and can be significant even for protons or
deuterons, at temperatures
of interest such as displayed in~Fig.~\ref{rankhu}.

The relativistic enhancement of the collective effect with
thermal agitation, suggests the possibility of a light-particle
distribution that is concave up on account of relativity,
when
the collective motion is present, and a logarithm of
the distribution is plotted vs energy.  On~the~other hand,
a distribution of collective velocities peaked at
finite velocity
values might generate a particle spectrum at low temperatures
or for large particle mass, that is concave down.  The
particle momentum distribution is given by
\begin{equation}
{d^3 N_X \over d p^3} = {g_X \over (2 \pi)^3} \int d {\bf r} \,
f_X
\end{equation}
and the slope of the distribution is
\begin{equation}
{1 \over T_{eff}} = - {\partial \over \partial E_X} \log \left(
{d^3 N_X \over d p^3} \right) = {1 \over T} \left( \langle
\gamma \rangle _{\bf p} - {E_X \over p} \langle \gamma {\bf n}
\cdot {\bf v} \rangle _{\bf p} \right) ,
\end{equation}
where ${\bf n}$ is a unit vector in the direction of ${\bf p}$
and
\begin{equation}
\langle \cdot \rangle _{\bf p} =
\int d{\bf r} f_X \,(\cdot) \left/ \int d{\bf r} f_X  \right. .
\label{avp}
\end{equation}
The slope at $p = 0$ becomes
\begin{eqnarray}
{1 \over T_{eff}} & = & {1 \over T} \left( \langle
\gamma \rangle _{p=0} - {m_X \over T} \langle \gamma ({\bf n}
\cdot {\bf v})^2 \rangle _{p=0} \right) \nonumber \\ & \approx
& {1 \over T} \left( 1 + {\langle
v^2 \rangle _{p=0} \over 2} - {m_X  \langle
({\bf n} \cdot {\bf v})^2 \rangle _{p=0} \over T} \right) ,
\label{slop0}
\end{eqnarray}
where the approximation leading to the last expression is for
low
$v$. Note that the slope (\ref{slop0}) may be negative, which
corresponds to a situation when the spectrum has a maximum
at a~finite $p$.
 For a high momentum $p$, the dominant contribution to the
integrals in (\ref{avp}) would come from a velocity directed
along ${\bf p}$ to an extent possible given a velocity
distribution,
on account of the strong dependence
of (\ref{feq}) on ${\bf v}$ for large $p$,
of a magnitude $v_m < p/E_X$ assuming a rapidly
falling velocity-distribution.
The slope may be then estimated as
\begin{eqnarray}
{1 \over T_{eff}} & \approx & {\gamma_m \over T } \left( 1 -
{v_m E_x \over p} \right) \nonumber \\ & \approx& {1 \over T}
\sqrt{{1 -
v_m \over 1 + v_m}} \approx {1 \over T} \left( 1 - v_m + {v_m^2
\over 2} \right)
\label{slopi}
\end{eqnarray}
where the second approximation is for $E_X \sim p$ and the last
approximation is for small~$v_m$.  If~(\ref{slop0})~is larger
than (\ref{slopi})
then the the spectrum following from (\ref{feq}) is necessarily
concave up within a~certain
range of energies; if (\ref{slopi}) is larger then the spectrum
is concave down (see also~\cite{sch93}).  For example, if
collective expansion has a cylindrical symmetry and
collective
speeds are sharply concentrated around a single value $v$,
then the condition for the spectrum at
90$^{\circ}$ with respect to the symmetry axis to be concave
down becomes
\begin{equation}
m_X > {2 T \over v} .
\label{mXT}
\end{equation}
Clearly, we may have a situation when a heavy-particle
spectrum is concave down, while a~light-particle spectrum is
concave up.  For a spherically symmetric expansion, the factor
of~2 on the r.h.s. of (\ref{mXT}) gets replaced by 3.
Explicit form of a distribution for a cylindrically
symmetric
expansion with a single value of $v$ is (see also~\cite{sch93})
\begin{equation}
{d^3 N_X \over d p^3} = {\cal N} \,
\exp \left( - {\gamma E_X\over
T}\right)
\,
I_0 \left( {\gamma p v \sin \theta \over T } \right)
\label{2d}
,
\end{equation}
where ${\cal N}$ is a norm, $I_0$ - a modified Bessel function,
and $\theta$ - an angle relative to the symmetry axis.  For a
spherically symmetric expansion with a single value of $v$, the
distribution in momentum is
\begin{equation}
{d^3 N_X \over d p^3} = {\cal N}   \,
\exp \left( -
{\gamma E_X \over T} \right)
\,
\mbox{sinh} \left( {\gamma p v \over T } \right)
\left/ \left( {\gamma p v \over T } \right) \right.
{}.
\label{3d}
\end{equation}
If a chemical potential $\mu_X$ is uniform throughout
a volume $V$, then the normalization constant in (\ref{2d}) and
(\ref{3d}) is ${\cal N} =  g_X V \exp (\mu_X/T)/(2 \pi)^3$.
For smooth distributions of collective velocities, the
sufficiently
heavy particles (and in principle also possibly the sufficiently
energetic) probe directly the collective velocity distribution
\begin{eqnarray}
{d^3 N_X \over d p^3} & = & {g_X \over (2 \pi)^3} \int d {\bf
r} \, f_X \nonumber \\
& \approx & {g_X \over (2 \pi)^3} \int d {\bf r} \, \exp \left[
{\mu_X - m \over T} \right] \exp \left[ - {E_X^4 \over m_X^3 T}
\left({\bf v}({\bf r}) - {{\bf p} \over E_X} \right)^2 \right] \nonumber
\\
& \approx &  {1 \over \gamma^6} \left( {T \over \pi m_X } \right)^{3/2} \, {g_X
\over (2 \pi)^3} \int d {\bf r} \, \exp \left[
{\mu_X - m \over T} \right] \delta \left(
{{\bf p} \over E_X} - {\bf v}({\bf r}) \right) ,
\label{heavy}
\end{eqnarray}
where the argument in the exponential in $f_X$ was expanded
around the maximum, as~a~function of ${\bf v}$, and
it was assumed that the velocity distribution, nominally
defined as
\begin{eqnarray}
{\partial^3 {\cal P}_X \over \partial v^3}
& = & \int d{\bf r} d{\bf p} \, f_X \,
\delta({\bf v}-
{\bf v}({\bf r}) ) \left/ \int d{\bf r} d{\bf p} \, f_X
\right. \nonumber \\
& = & \gamma {\cal N}'
 \int d {\bf r} \, \exp \left[
{\mu_X - m \over T} \right] \delta \left( {\bf v} -
{\bf v}({\bf r}) \right) ,
\end{eqnarray}
 changes
slowly
over the range in $v$ of a magnitude $\gamma ^{-2}
(T/m)^{1/2}$.

Besides collective motion, the spectra of different
particles may be affected, at low momenta, by resonance
decays and statistical effects.  Following (\ref{feqs}),
corrections for statistics in~the~case of Eqs.
(\ref{2d}) and (\ref{3d})  amount
to adding or subtracting from the r.h.s. terms corresponding to
lower temperatures.  In order to assess a relative
importance of different effects, it will be necessary to
understand the dynamics of production.  With pions not
present in the initial state, an important pion observable that
may
be affected by collective expansion, besides spectra, is
multiplicity.

\subsection{Dynamics of Production}

Production of pions in central La + La reactions at
800 MeV/nucleon will be considered.  At different values of
beam energy, pion multiplicities have been measured in La +
La reactions~\cite{sto82}.  At 800 MeV/nucleon, both
pion and proton spectra were determined at c.m.~90$^{\circ}$
in the very central reactions~\cite{hay88}.
Using information~in~\cite{hay88}, these spectra may be
conveniently normalized for comparisons with calculations.

Figure \ref{profla} shows the evolution of baryon-density
profiles, along and perpendicular to the beam axis, in the
La + La reaction at $b = 0$.  The system goes through
similar stages as Au~+~Au in Fig.~\ref{profau}.  Figure
\ref{densla} shows further baryon density at the center
of the La + La system, net participant collective energy
per nucleon
calculated as in Eq. (\ref{Ecol}), and total entropy as
functions of time.  The collective energy saturates at about 76
MeV/nucleon, out of which $\sim 56$ MeV/nucleon is associated
with
transverse directions and $\sim 20$ MeV/nucleon with longitudinal.
The difference in the energy per degree of freedom for
different
directions is not as large as in the 400 MeV/nucleon Au + Au
reaction due to a smaller system size, combined with a larger
Lorentz contraction and a reduced stopping at the~higher beam
energy.

One
deficiency of the present model is that composites
produced
during the expansion do not participate in pion absorption
and production, while in reality they would.  To~remedy this,
some details relevant to
pion production, such as variation with time of particle
collision-rates and $\Delta$
decay rates in Fig.~\ref{rates},
are presented for a calculation with production of
composites switched off.  However, final pion spectra are given
for a calculation with composite production switched on, as
composites
raise the freeze-out temperature,
reducing to a certain degree the slopes of spectra.  The
final
pion multiplicity does not depend in practise on composites
as will become apparent later.

In Fig.~\ref{rates} it is seen
that the rise of $\Delta$-decay rate is somewhat
delayed relative to the rise of $\Delta$-production rate in
NN collisions.  This is due to a~finite $\Delta$ lifetime (see
also~\cite{cug82}).  The
rise of the resonance formation rate in N$\pi$ interactions
follows very closely the rise of $\Delta$ decays, on account of
a sizeable average N$\pi$ cross section.  With the exception of
the $\Delta$-production rate in NN interactions, that maximizes
earlier, all rates
maximize at $t \sim 13.5$~fm/c, when the shock waves in
lanthanum nuclei reach the vacuum, see Fig.~\ref{profla}, and
the system as a~whole becomes excited and compressed.  A~later
decrease of rates in Fig.~\ref{profla} is associated
with expansion.

As pions (within the model) can only originate from
nucleon-resonance decays, and~pions can only be absorbed into
nucleon resonances,
the final pion number is equal to the difference in the total
number of resonance decays and the number of resonance
formations in $\pi$N
interactions, i.e.~the~difference in areas under the curves
in the bottom panel of Fig.~\ref{rates}.  (The N$^*$'s
of which the number at 800~MeV/nucleon at any one time
never exceeds 4\% of the total number of resonances,
may be ignored in this discussion.) The rather small
difference between the rates compared to the rates themselves
at most times suggests a~good kinetic
equilibrium of pions and deltas with nucleons during the
reaction.  As, clearly, the
number of deltas contributing a~final pion is identical
with the net
number of deltas produced in nucleon interactions, the
final pion number further coincides with the difference in
areas under the curves in the center panel of Fig.~\ref{rates}.
If one considers a chain starting with a delta production in
an NN interaction followed by a delta decay with pion
emission, followed by an absorption of this pion into a~delta
and so on, then such a
chain terminates either with a~pion in the final state, or with
a~delta absorbed in an interaction with a nucleon.  The~average
number of pion absorptions within any such chain is equal to
the ratio of an~area under the
dashed line in the bottom panel in Fig.~\ref{rates}, to an
area under the solid
line in the center panel, i.e.  $\sim$1.5.  If only chains
terminating with a final pion are considered, then the number
of absorptions is found to exceed 2 in the simulation.  One
can say
that a pion in the final state of the system has been, on the
average,
absorbed more than twice and reemitted before reaching
the vacuum (see also recent papers~\cite{bas93,bas94}).

Pion multiplicity measured in central collisions is lower by
nearly a factor of 2 than anticipated in the fireball model which
assumes a complete global equilibration~\cite{sto82}.  A naive
expectation
might be that within the reaction the time is too short to
allow
for a production of the number of pions and
deltas expected in a global equilibrium.
(The deltas left out towards the end of reaction would
contribute to a final pion number.)
Figure \ref{deltap} shows
the pion and delta number divided by
total nucleon
number, as a function of time in the La~+~La reaction at 800
MeV/nucleon.
The number reaches a maximum right after the
whole matter has entered the hot compressed region.
The ratio of the
global-equilibrium number of pions and deltas to baryon number
at this time, $t \sim14$ fm/c, is
indicated by the upper of two horizontal dashed lines in
Fig.~\ref{deltap}.  The ratio is obtained using
the equilibrium densities
\begin{eqnarray*}
\rho _{\pi} = 3 \int {d{\bf p} \over (2 \pi)^3 } \, {1 \over \exp
(E_{\pi}/T) - 1},  \hspace{3em}
\rho _{N} = 4 \int {d{\bf p} \over (2 \pi)^3 } \, {1 \over \exp
((E_{N} - \mu_N)/T) + 1},  \nonumber \\
\end{eqnarray*}
\begin{equation}
\rho_{\Delta} = 16 \int {d{\bf p} \over (2 \pi)^3 } \int {dm
\over 2 \pi}  \, {\cal A}_{\Delta} \, {1 \over \exp ((E_{\Delta}
- \mu_N)/T) + 1} , \label{rheq}
\end{equation}
where temperature and chemical potential are adjusted to
reproduce the representative baryon density, $\rho =
2.25 \rho_0$ at a respective time (cf. Fig.~\ref{profla}), and
the energy per nucleon from simulation, excluding potential
energy,
of 1133 MeV/nucleon.  (Isospin asymmetry is ignored for
simplicity.)  The equilibrium
value, indicated by the upper dashed line in Fig.~\ref{deltap},
is not reached, what appears to support an expectation that the
time is too short to produce an adequate number of pions and
deltas.  In fact, as~the~system expands and baryon density
drops, the global equilibrium value increases more and
likewise the apparent deficit of pions and deltas.
Towards the freeze-out, however, the number of pions and
deltas behaves in an~unexpected manner.
Instead of
straining towards the equilibrium value and settling once interactions
stop, this number declines.
The~decline
is seen
in many other microscopic calculations of collisions,
e.g.~in~\cite{sto86,cug82,wol90,fuc93}.  (Incidentally,
if~a~ratio of pion
and delta density to baryon density is investigated locally, then
the decline becomes even more pronounced; the~total number
represents the system as a whole with different parts going
through different stages.)  The~rise and decline in the number
of
pions and deltas could be understood if somehow the
number of pions and deltas were first below and later
above the equilibrium  number.
The clue to an understanding of this situation is in
collective motion ignored till now.

Collective
motion within the system takes away energy available locally
for particle production.  Right after the whole matter
in the La + La reaction has entered the hot region, at $t \sim
14$ fm/c, the collective energy within this system, as seen
in~Fig.~\ref{densla}, is~$\sim 30$~MeV/nucleon.  If this energy is
subtracted from the available
energy and Eqs.~(\ref{rheq}) are used as before, then the
equilibrium value for the ratio
of pions and deltas to baryons is obtained which is indicated
by the lower of two horizontal dashed lines in~Fig.~\ref{deltap}.
It~is apparent in this figure that the system approximately
reaches an  equilibrium with
regard to the pion and delta number given the particular value
of collective energy.  As~the~system expands,
though, more and more energy turns into collective.
In~Fig.~\ref{entro} it is seen that if expansion were isentropic then
the
number of pions and deltas should decrease.  (At low
temperature, in~an~isentropic expansion the pion-to-baryon
ratio behaves as $\rho_{\pi}/\rho \propto \exp(-m_{\pi}
\rho_1^{2/3}/T_1 \rho^{2/3})$.)  In~Fig.~\ref{entro}, the state
of the system
at $t = 14$ fm/c is crudely (as in fact different parts of the
system go through different stages) represented with crosses.
To assess the
situation further, the equilibrium number of pions and deltas
is calculated at different times by evaluating the equilibrium
density of particles throughout the space in local frames,
given energy per baryon and baryon density in such a frame, and
integrating the particle density over whole space.  Results are
represented with
dots in~Fig.~\ref{deltap}.  The~obtained local-equilibrium
number of pions and deltas is seen to
 increase when the mass of hot matter increases,
and to decrease rapidly afterwards.  The~actual number of the
particles closely follows the equilibrium number during
compression.
The local-equilibrium number crosses the actual number of pions
and deltas at
about the maximum of the latter number (as~should be expected).
While the number of pions and deltas decreases during
the expansion, it freezes out, nonetheless, at a relatively
high value.  This is because a~decrease of nucleon density
in expansion effectively shuts off the processes $N+N
\leftrightarrow N + \Delta$,
see~Fig.~\ref{rates}.
The~processes of pion absorption, with a~larger cross
section, and the $\Delta$ decays continue somewhat longer,
preserving the~thermal but not the~chemical equilibrium of the
pionic sector with
nucleons.  Portion of the chemical equilibrium path, with
nonadiabaticity
taken into account, is sketched with dashed lines
in~Fig.~\ref{entro}.  At~a~freeze-out, at a density $\rho \sim
0.5 \,\rho_0$, the~ratio of pions and deltas to baryons
is larger by $\sim (1.4-1.8)$ than in the state of chemical
equilibrium.

It has been conjectured in the past~\cite{sto82} that pion
multiplicity in the central collisions is low, because part of
the energy in the center of mass is used up in compressing
matter,
and, correspondingly, is not available for production.  Present
results show
that the role of energy not available for production
within a heavy system is played
by the collective energy.  In~fact,
potential energy at the time when
pion and delta number maximizes in the La + La
reaction simulation, differs
from the potential energy in a ground state by
a~mere~3~MeV/nucleon.  (Momentum-dependent interactions might,
though, have an effect on pion number.)
With~regard to another expectation, that the time in a reaction
is too short for
producing an~equilibrium number of pions and deltas, it is seen
that the number of pions and deltas actually freezes out not
below, but above an equilibrium value.

\subsection{Features of Single-Particle Observables}

The left panel in Fig.~\ref{spectra} shows calculated
momentum distributions of different particles~(filled~symbols)
within a c.m. polar-angle range of
60-120$^{\circ}$ in the 800 MeV/nucleon La~+~La reaction at~$b
=
1$~fm, together with momentum distributions of protons
and negative pions in the very central reactions deduced from
experimental results of Ref.~\cite{hay88} (open symbols).
The~cross section for central
reactions investigated in Ref.~\cite{hay88}, needed for the
normalization of distributions was estimated at 86 mb, using
the
integrated $\pi ^-$ yield stated~in~\cite{hay88} and
$\pi$~multiplicities
from~Ref.~\cite{sto82}.  The experimental $\pi ^ -$
distribution agrees very well with the calculated distribution
at all energies, and the proton distribution -- up to an energy
of $\sim 450$~MeV in the c.m.s.  The pion distributions may be
well approximated with straight lines (in the logarithmic
plot), while baryon distributions are concave down.  The~slopes
significantly decrease with increasing particle mass.
If the calculated slopes were to be interpreted in terms of
temperature, quite disparate values would be obtained, of $70$,
$\sim 110$, $\sim 145$, and $\sim 175$ MeV, for pions, protons,
deuterons, and helions, respectively.

For the sake of understanding the momentum distributions from
simulations,
the left panel in Fig.~\ref{spectra}, shows additionally
baryon and neutral-pion distributions
from an instantaneous freeze-out model (solid lines).
In the case of an expansion with a single value
of velocity,
the~asymptotic collective energy in Fig.~\ref{densla}
would correspond to $v \approx 0.41 c$.
In the case of a uniform velocity distribution, adopted
within the model for the display in Fig.~\ref{spectra},
the~collective energy
corresponds to a cut-off velocity-value of $v_c \approx 0.51c$.
The mean proton energy at a freeze-out of 196 MeV from
simulation, corresponds then to a local temperature of $T
\approx 55$~MeV.  The baryon momentum distributions within the
model take the form
\begin{equation}
\label{dnX}
{d^3 N_X \over d p^3} = {g_X V  \over (2 \pi)^3 } \, \exp
\left(
{\mu_X \over T}\right)\, {3 \over v_c^3} \int_0^{v_c} dv \,
v^2 \,
\exp \left( -
{\gamma E_X \over T} \right)
\,
\mbox{sinh} \left( {\gamma p v \over T } \right)
\left/ \left( {\gamma p v \over T } \right) \right. ,
\end{equation}
cf. (\ref{3d}).
For pions the distribution is (see also~\cite{bar81,hah88})
\begin{eqnarray}
{d^3 N_{\pi^0} \over d p_{\pi}^3} & = &
{ V   \over (2 \pi)^3
} \, \exp \left( {\mu_{\pi^0} \over T} \right) \, {3 \over
v_c^3} \int_0^{v_c}  dv \,   v^2
\Bigg\{ \exp \left( - { \gamma E_{\pi} \over T}
\right) \mbox{sinh} \left( {\gamma p_{\pi} v \over T } \right)
\left/ \left( {\gamma p_{\pi} v \over T } \right) \right. \cr
& & +
 \exp \left( {\mu_{\pi^0} - 2 \gamma E_{\pi} \over T}
\right) \mbox{sinh} \left( {2 \gamma p_{\pi} v \over T } \right)
\left/ \left( {2 \gamma p_{\pi} v \over T } \right) \right. \cr
& & + {16 \over 3} \exp \left( {\mu_{N} \over T} \right) {1 \over 2 p_{\pi}
E_{\pi} } \int {dm \over 2 \pi} \, {\cal A}_{\Delta} \, {
m \over p_{\pi}^*} \cr & & \times \int_{p_{\Delta
1}}^{p_{\Delta 2}} d p_{\Delta} \,  p_{\Delta}
\exp \left( -
{\gamma E_{\Delta} \over T} \right)
\,
\mbox{sinh} \left( {\gamma p_{\Delta} v \over T } \right)
\left/ \left( {\gamma p_{\Delta} v \over T } \right) \right.
\Bigg\}
\end{eqnarray}
where the second and third term inside braces account,
respectively, for the effects of Bose statistics in lowest
order, and for the decays of $\Delta$ resonances left-out at
a freeze-out.
The~minimum and maximum momenta of a $\Delta$ whose decay might
yield a pion with a~momentum~$p_{\pi}$, are
 $(p_{\Delta})_{1,2} =
(m/m_{\pi}^2)|E_{\pi}^* p_{\pi} \mp E_{\pi} p_{\pi}^*| $.  The
subscript '$^*$' indicates
pion momentum and energy in a $\Delta$-frame.  Pions and
deltas are assumed to be in a thermal but not in a chemical
equilibrium with the rest of the system; the chosen value of
a freeze-out density is $\rho = 0.6 \rho_0$.

It is quite apparent in Fig.~\ref{spectra} that the
instantaneous freeze-out model (with parameters taken from
dynamics) describes,
in~a~semi-quantitative manner, the shapes of spectra from
simulations.  The spectra of the lightest particles, pions and
protons, with largest thermal dispersion of velocities, are, in
fact, equally well described within a model with a single
collective expansion velocity.  Proton mass (and~cluster
masses) satisfies $m > 3T/v$ and the spectrum is concave down.
For~pion mass the opposite inequality is true.  Long-- and
short--dashed
lines in Fig.~\ref{spectra} show, respectively, contributions
from free pions and from $\Delta$ decays within the
instantaneous freeze-out model.  The free-pion contribution is
concave up, but in the overall distribution this concavity is
partially filled up with pions from decays.  Despite the spread
of the $\Delta$ resonance in mass, a~major resonance
contribution comes
here (and at other temperatures of interest) from masses in the
vicinity
of resonance peak; the~ratio of resonances to pions is
$\rho_{\Delta}/\rho_{\pi} \propto \rho_N \,
\exp(-(m_{\Delta}-m_N - m_{\pi})/T) /T^{3/2}$.

Of a general interest is the degree of Bose enhancement or
the proximity to a condensation for low-momentum pions.  The
excessive number of neutral pions at a freeze-out corresponds
to a finite chemical potential $\mu_{\pi^0} \sim T \log
(N_{\pi^0}/N_{\pi^0}^{eq}) \sim 25$ MeV.  For negative pions,
the isospin asymmetry within the system corresponds to an
additional enhancement of the potential for negative pions, and
a depletion for positive pions, $\mu_{\pi^-} - \mu_{\pi^0} =
\mu_{\pi^0} - \mu_{\pi^+} \sim T \log(N/Z) \sim 20$~MeV.
Overall, the potential at a freeze-out for negative pions may
be as large as~$\sim 45$~MeV, corresponding to a low-momentum
enhancement of $\exp((\mu_{\pi^-} - m_{\pi^-})/T) \sim 18 \%$.
Possible enhancement for neutral pions is 13\%.  Beyond
that,
the Coulomb field in the final state enhances
distribution of negative pions at low momentum and depletes
that of positive
pions and other particles.  Lower limit on the magnitude of the
effect
on single-charged low-momentum particles follows from
the formula obtained under an~assumption of a~dominating
collective motion,
$0.7$~MeV$\cdot$fm$\, \rho^{1/3} Z/A^{1/3}
\langle E^{kin} \rangle \sim 5 \%$, where $Z$ and $A$ are total
charge and mass numbers, respectively, $\langle E^{kin}
\rangle$ is mean kinetic energy of particle species in
question,
and~$\rho$~is freeze-out density.  Inspection of spectra
reveals that the actual effect of the Coulomb field in the
La~+~La reaction is of the order of 30\%.

Collective motion is generally characterized by a correlation
between particle spatial positions and their momenta.
In order to demonstrate explicitly that strong features
in the calculated
distributions, particularly slope-differences between
different particle species, are~associated with the~collective
expansion, calculations are further repeated with an
interchange
of particle positions during evolution.  Interchanges are made
within same species, when collisions
are still frequent.
(Particle momenta are not altered.)
The  procedure leaves
density in configuration space and momentum distributions
intact.  Correspondingly, kinetic and potential energies stay the same.
However, correlations between particle momenta and
positions get destroyed.  The final momentum distributions from
calculations with the position interchange are shown in the
right panel of~Fig.~\ref{spectra}.  Straight parallel lines are
drawn to guide the
eye.  The distributions from the calculations turn out largely
exponential with nearly the same slopes (some collective motion
still develops due to the collisions, cf.~Fig.~\ref{densla}).
There is enough time within a reaction for a complete
thermalization.

Figure \ref{pimul} shows measured pion multiplicities in
central symmetric reactions as a function of beam
energy~\cite{sto82},
together with multiplicities from standard calculations of
central La~+~La reactions (solid line) and from calculations
with positions interchanged (dashed line).  The standard
calculations,
with the~collective motion developed during evolution, describe
reasonably~the~data.  The calculations with the position
interchange and the collective motion extinguished, on the
other hand, lead to multiplicities larger by nearly a~factor
of~2 (essentially
such as expected within the fireball model), in a dramatic
disagreement with data.

Despite of the fact that the measured pion multiplicities point
strongly towards the collective motion, in order to better
assess this motion, it would be clearly highly
desirable to have {\em normalized pion and proton}
momentum
distributions from central collisions of heavy nuclei,
extending
over a wider kinetic-energy range than in~Fig.~\ref{spectra},
possibly together with cluster distributions.  Heavy systems
are preferred over light systems~\cite{sie79}.  In
peripheral
collisions of light systems, differences between pion and
proton spectra may exceed differences arising from
collective-energy
considerations, on account of energy conservation in
elementary
collisions, with kinetic energy depleted due to the
production of pions.  In~principle,
with~a~decrease of impact
parameters, the~relative importance of
conservation in individual
collisions would decrease, while the importance of collective
motion should increase.  To~test an~interpretation in terms of
collective energy, features of
the cluster distributions could be examined~\cite{man82}.
In~lightest systems, though, such an~interpretation would be
always hampered by the lack
of a complete stopping even in most central collisions, at~beam
energies at which pions are abundant.

The actual distribution of collective velocities for baryons in
the vicinity of last collisions from the simulation of the
800~MeV/nucleon La~+~La
reaction is shown in~Fig.~\ref{coldis}.  This distribution is
not peaked at any large value of velocity (although $v^2(d^3
{\cal P}/dv^3)$ is peaked, however broadly, in the vicinity of
$v \sim  0.4 c$).  Tests indicate that
the high-velocity tail is associated with low temperatures.
High temperatures, on the other hand, are particularly
associated with the low-velocity portion of the distribution.
Parametrization of particle spectra, such as from experiment,
in terms of a single freeze-out temperature for any particle,
will be always approximate.  Also an approximation can
generally
arise from the use of same temperature and same velocity for
different particles, as e.g. pions may be emitted earlier when
temperatures are slightly higher and the collective energy
lower,
than when light baryons are emitted, and differences generally
develop in the emission of different baryons, cf. Sec.
\ref{frapro}.  In any case, though, the use of a
uniform
distribution in collective velocity in parametrizing particle
distributions would be more sensible than
the use of a~single velocity value.  One possibility within an
experiment is the use of a velocity distribution deduced from
heavier
fragments, of which distributions should approach that of the
collective velocity, with increasing fragment mass, cf. Eq.
(\ref{heavy}).
The use of a single velocity value and of a~temperature treated
as adjustable parameters in describing measured spectra of
different
particles, on the other hand, is expected to give absurd
results.  Reasonable
temperatures would lead to excessively sharply-peaked
distributions of heavy fragments.  To make the fits acceptable,
temperature parameters would then need to be raised making it
appear that heavier fragments came from hotter regions
than light fragments.

\subsection{System Measured by Brockmann et al.}

Before moving on to other issues, let me discuss
data~\cite{bro84},
that first stirred up interest in shapes of transverse spectra
of pions from
central reactions, from the Ar + KCl system at~1.8~GeV/nucleon,
and further aspects of transport models.
In~\cite{bro84}~experimenters attempted to fit their data
with
a single exponential and concluded that this was not possible.
Moreover, they compared results to those from a cascade
model~\cite{cug82}.  Due to the low mass of the system and
high beam energy, and corresponding lack of stopping, the
spectra from transport simulations are quite sensitive to the
assumptions
on elementary collisions; the experimenters chose a~cross
section for $N + N \rightarrow N + \Delta$ processes that
was exceedingly forward-peaked.
Cascade
results were not laid over data, but rather fitted with an
exponential.  The~conclusion was that the cascade model
cannot explain the {\em high}-momentum portion of a pion
spectrum.

The data from~\cite{bro84} are shown in two panels of
Fig.~\ref{ar18} (open circles).  Insets show an~asymmetry
coefficient $a$
obtained by fitting the double-differential distribution
\begin{equation}
{d^2
N_{\pi^-} \over dE \, d \cos \theta } \propto (1 + a(E) \cos
^2 \theta ) .
\label{asym}
\end{equation}
The right panel shows further results from the cascade
model~\cite{cug82} following
assumptions~in~\cite{bro84}~(squares).
It is seen that, indeed, the high-momentum tail of the measured
pion distribution is underpredicted.  Further, the asymmetry
coefficient from the model is far too high at high pion kinetic
energies.  Discrepancy in this coefficient reveals part of the
problem with the high-momentum tail.
The left panel of Fig.~\ref{ar18} displays, in addition to
data~\cite{bro84}, results
from the present model~\cite{dan91} with an angular dependence
of the delta-production cross-section that was fitted to
elementary data
in~Ref.~\cite{wol92}~(filled~circles).  The~model describes
well
the high-momentum tail of the pion spectrum, and it describes
much
better the asymmetry coefficient than the cascade model.
This is associated both
with the shape of elementary delta-production cross-section in
the model
and with respecting of the time-reversal invariance in
absorption
cross-sections (cf. Sec. \ref{Model}; high-energy pions are
excessively absorbed in the original cascade model compared
to the low-energy pions).  For~illustrative
purposes (the involved assumption is too extreme), calculations
were repeated within a~version of the cascade model where
$\Delta$-production cross-section was isotropic;
the~results are shown in the right panel of Fig.~\ref{ar18}
(diamonds).  The description of the high-momentum
tail and asymmetry coefficient is improved over the
first cascade calculation.

Overlaying the calculations and data reveals
a puzzle concerning the {\em low}-momentum part of pion
spectrum.  Already the cascade model
underpredicts data~\cite{bro84} in
this region and correcting the pion absorption cross-section,
cf. Eq. (\ref{piNcs}) and associated text, increases
discrepancy.  Measured spectrum appears to have a low-momentum
peak as though
superimposed onto the spectrum such as from the calculation, of
a similar
height as the latter, see left panel in~Fig.~\ref{ar18},
apparently pointing
towards physics outside of the transport model.~\footnote{
Elsewhere, though, results well exceeding data in the
low-momentum region were obtained \protect\cite{bao91},
supposedly on
the account of the nuclear mean field and Pauli principle.
Here, little effect of these factors on spectra is found (as is
apparent in~Fig.~\protect\ref{ar18}) in accordance with the
intuition, given the small
system size, the high beam energy, and the momentum region.  No
attempts are
made here to interpret unnormalized data in literature.}

\subsection{Shadowing}

As impact parameter is increased in symmetric collisions at
high beam energies, regions with spectator matter develop and
grow in size.  In strongly asymmetric systems, such as
investigated in~\cite{gos89}, the spectator
matter is always present.  The matter can shadow pions emitted
from the hot participant region.

In Fig.~\ref{pxa} with mean momentum-components within the
reaction plane of particles emitted from
1 GeV/nucleon Au + Au reactions, it is seen that, at low impact
parameters $\sim 3$~fm, pions moving forward and backward in
the c.m.s. have  momentum components of same sign as
baryons.
As the pion angular-distribution is
forward-backward peaked at $b = 0$, cf.~Fig.~\ref{polau}, the average
deflection such as for
baryons, within the reaction plane, must be associated with the
weak discontinuity.  The pion momentum components per mass are
considerably lower than the baryon components,
see~Fig.~\ref{pxa}, and pion velocity components even more so,
by~a~factor of $\sim 5$ in the forward
and backward rapidity regions, than the baryon components.
At higher impact parameters ($\sim 7$ fm) in the reaction,
pions have momentum components  of an opposite
sign than baryons,
cf.~Fig.~\ref{pxa}.
These components increase more than twofold if
analysis is limited to pions emitted early, within the
first~16~fm/c of the reaction.  (The~latter pions constitute about 1/3
of all.)

When
pion emission is examined at a 90$^{\circ}$ c.m. polar-angle,
it is found essentially isotropic in the azimuthal
angle at low impact
parameters such as $b \sim 3$ fm.  However,
at
higher impact parameters, such as $b \sim 7$ fm,
this emission is
found
peaked at 90$^{\circ}$ with respect to the reaction plane.  The
peaking is enhanced when analysis is limited to pions with high
c.m. momenta, $p > 200$ MeV/c, see~Fig.~\ref{dNfi} ($\Theta =
0^{\circ}$).  The latter peaking was
observed experimentally in events corresponding
to intermediate impact parameters~\cite{bri93,ven93}.

Figure \ref{denspi} displays contour plots of baryon density
(solid lines) within the reaction plane, in~the~1~GeV/nucleon
Au + Au reaction at $b = 7$ fm, at a time $t = 15$ fm/c.
At the impact parameter equal
to the half of the maximum possible value, the spectator
regions are well developed. The
spectator matter begins to detach from participant matter
at the time at which the densities are displayed.
Shocks have started out in the particular reaction at an~angle
of inclination
relative to the beam axis of $\sim 65^{\circ}$, cf.
Eq. (\ref{ealpha}), and have
rotated down to lower angles $\sim 55^{\circ}$,
compare~Fig.~\ref{tang}.  The last of what has become the
participant matter has dived into the shocked region at a time
$t \sim 10$ fm/c.

Figure \ref{denspi} displays further contour plots of pion and
delta density within the reaction plane.  It is
apparent in that figure that
the spectator matter limits pion emission.  An analysis of the
emission in different directions, within the simulation, shows
that the transmission coefficient for
pions through the center of spectator matter does not
exceed 5\% during the first 16 fm/c of the evolution.  With
not very strong sideward-flow effects exhibited by pions in the
absence or near absence of spectator matter at low impact
parameters (at this and at other beam energies), the evidently
strong shadowing at the higher impact parameters
must generate sizeable average momenta per mass
opposite, due to geometry, to the momenta of nucleons.

Early proton emission
is affected by shadowing, too.  In the proton case,
though,
the~shadowing does not lead to such dramatic effects as in the
pion case, because
baryons exhibit generally stronger
dynamic effects and because baryon emission is overall
somewhat
delayed relative to pion emission.  Thus, the mean momenta
within
the reaction plane of protons decoupling from the
1~GeV/nucleon $b = 7$~fm Au~+~Au,
within the first~16~fm/c, are nearly zero.  These protons
constitute a smaller fraction of particles emitted at all times
than pions (about 10\% as compared to 1/3 for pions; pions get
ahead of the bulk of the matter on account of larger
thermal velocities).

While the effect of shadowing on the mean momenta is opposite
to that of dynamics,
both type of effects act to suppress the emission
within the reaction plane at midrapidity as compared to the
emission
out of the reaction plane.  Figure \ref{dNfi} with
calculational results illustrates the fact that the issue as to
which of the above two effects dominates in generating an
azimuthal
pattern at midrapidity, may be resolved at a qualitative level
within an~experiment.
Thus, one needs to examine an
emission pattern
within a plane $p^{z'} = 0$ where axis $p^{z'}$ points
at an angle $\Theta$ measured counterclockwise within the
$p^x$-$p^z$ plane, from the beam.  When shadowing
dominates, an azimuthal anisotropy should persist or increase
compared to that at $\Theta = 0$ (within a plane
perpendicular to the beam), when axis $p^{z'}$ is rotated to
negative
$\Theta$-values, making the plane of analysis
more aligned with spectator pieces when latter
are close to the participant matter, at a time when many
particles are emitted,
cf.~Figs.~\ref{denspi}~and~\ref{contours}.  On the
other hand, an anisotropy solely due to shadowing should
decrease or disappear when axis $p^{z'}$ is rotated to positive
$\Theta$-angles
(case of pions in~Fig.~\ref{dNfi}).  When anisotropy is
primarily associated with a stronger collective motion in
the directions parallel to shock-front surfaces, in particular
in the direction out-of-the-reaction-plane, in~comparison to
the direction perpendicular to shock surfaces,
then the anisotropy
within the plane $p^{z'}=0$ should increase when axis
$p^{z'}$ is rotated to positive $\Theta$-angles (case of
protons in Fig.~\ref{dNfi} and baryons in
Refs.~\cite{gut89,gut90}).

With regard to the transverse-momentum dependence of
azimuthal anisotropies~\cite{bri93,ven93}, any
anisotropy must disappear
when $p^{\perp} \rightarrow 0$ (but see~\cite{bas93}),
provided that particle momentum distribution,
denoted below as
\begin{eqnarray*}
{\cal N}({\bf p}^{\perp}) \equiv {d^3 N_X \over
dp^3} ({\bf p}^{\perp},p^{z}) ,
\end{eqnarray*}
is smooth at $p^{\perp} = 0$.
Expansion of the distribution
up to the second order in transverse momenta yields
\begin{eqnarray}
{\cal N}({\bf p}^{\perp}) & \simeq & {\cal N}(0) + {1 \over 2}
\left[ \left. {\partial^2 {\cal N} \over \partial p^{x2} }
\right|_{p^{\perp} = 0} +
\left. {\partial^2 {\cal N} \over \partial p^{y2} }
\right|_{p^{\perp} = 0} \right] \, p^{\perp 2} +
\left. {\partial {\cal N} \over \partial p^{x} }
\right|_{p^{\perp} = 0} \, p^{\perp} \cos \phi
 \cr & &+ {1 \over 2}
\left[ \left. {\partial^2 {\cal N} \over \partial p^{x2} }
\right|_{p^{\perp} = 0} -
\left. {\partial^2 {\cal N} \over \partial p^{y2} }
\right|_{p^{\perp} = 0} \right] \, p^{\perp 2} \cos 2\phi ,
\label{Nexp}
\end{eqnarray}
where $\phi$ is azimuthal angle relative to the reaction plane
and $x$ and $y$ denote directions perpendicular to the beam
axis in-- and out--of--the--reaction--plane, respectively.
The coefficient in front of $\cos \phi$  vanishes at
midrapidity
in a symmetric system.  The coefficient in
front of $\cos 2\phi$ is proportional to
momentum squared.  (These features of the coefficients would be
retained
if third axis were rotated away from the beam.)  The azimuthal
asymmetry at midrapidity may be e.g. quantified with a ratio of
out-of-plane
to in-plane momentum distribution minus 1.  This~quantity is
proportional to the coefficient in front of $\cos 2 \phi$ in
Eq.~(\ref{Nexp}) and, correspondingly, vanishes quadratically
with momentum as $p^{\perp} \rightarrow 0$.  (That is also
true
for other measures of asymmetry such as $\langle \cos 2 \phi
\rangle$.)  In
the case of pions the asymmetry would grow with an increase of
pion
velocity at midrapidity, as~pions would move out to the sides
in a~reaction and encounter spectator matter in
the vicinity of the reaction plane.  At lower
impact parameters, when layers of spectator matter are thin and
semi-transparent to pions, some dependence of the asymmetry on
pion momentum can be expected, related to the
energy-dependence of pion-nucleon cross-sections.  At high
impact parameters in heavy systems, when layers of spectator
matter are thick and
intransparent to pions even for low pion-nucleon
cross-sections, the asymmetry should saturate once
pion velocities get close~to~$c$.  When this paper was under
completion, an extended study of pion shadowing by Li
has appeared~\cite{bao94}, with calculational results
exhibiting these features.  Data~\cite{bri93,ven93} are
suggestive of the
rise of asymmetry at high momenta, but~corresponding
errors
are large.  Other recent theoretic references on pion
shadowing in symmetric systems are~\cite{bas93,bas93a,bao93a}.

\section{Baryon Observables}
\label{frapro}
\subsection{Entropy and Fragment Yields}
Experimentally yields of different fragments are used to
determine entropy produced in reactions.  Specifically, at low
densities when effects of statistics and interactions between
fragments can be ignored,
thermodynamic entropy per baryon can be expressed as
\begin{equation}
{S_B \over A} = \sum_X {M_X \over A} \, \left( {5 \over 2} -
{\mu_X + B_X \over T} \right)
\approx {5 \over 2} \, {M \over A} -
{Z \over A} \, {\mu_p \over T} - {N \over A} \, {\mu_n \over T}
\label{SBA}
\end{equation}
where $M_X$ is multiplicity of fragment $X$, $B_X$ ($> 0$) is
fragment binding energy, and $\mu_X = Z_X \mu_p + N_X \mu_n$ is
chemical
potential.  The approximation in (\ref{SBA}) is for small
binding energies per nucleon in comparison to nucleon chemical
potentials $\mu_n$ and $\mu_p$.  The ratios of nucleon chemical
potentials to temperature can be obtained from fragment yield
ratios, thus e.g. $\mu_n / T = \log ( g_p m_p^{3/2} M_d/ g_d
m_d^{3/2} M_p) = \log (M_d/3 \sqrt{2} M_p)$, and
$\mu_p / T = \log ( \sqrt{2/3} M_h / M_d)$.  High entropy
values correspond to large negative values of $\mu_N /T$ and to
a rapid drop of yield with fragment mass.  In data
analysis~\cite{dos88a,kuh93}, theoretic models are used, such
as~QSM~\cite{sub81}, to extend the relation (\ref{SBA}) between
entropy and fragment yields to situations
when e.g. fragment phase-space overlap becomes of
a concern.

Entropy per nucleon determined from measured yields of
light fragments at large c.m. angles in the Au~+~Au reaction
at~400~MeV/nucleon~\cite{dos88a}, is shown
in the left panel of Fig.~\ref{enmu}
as a function of reduced participant-proton multiplicity.  The
right panel of Fig.~\ref{enmu} shows entropy per nucleon
associated
with fragments emitted at large angles~\cite{dan91} (circles),
determined from calculated fragment phase-space distributions,
as a function of $b$.  An~agreement is found with regard to
the value of entropy produced in the very central reactions and
with regard to the general
behavior of entropy with the reaction centrality at the
particular beam energy; see~further Ref.~\cite{dan91}.

In order to understand the production of entropy in collisions
it is convenient to consider first the case of $b = 0$.  In
that case, at high energies, most of the entropy is produced in
shock fronts, cf. Figs.~\ref{profau} and \ref{rankhu}.  Further
production takes place during expansion.  The~hydrodynamic
Navier-Stokes equations yield for the rate of change of entropy
per baryon with time~\cite{lan59}
\begin{equation}
{\partial \over \partial t} \left( {S \over A} \right) + {\bf
v} \cdot {\bf \nabla} \left( {S \over A} \right) = {1 \over
\rho T} \hbox{div} \left( \kappa {\bf \nabla} T \right) + {\eta
\over 2 \rho T} \left( {\partial v_i \over \partial r_k} +
{\partial v_k \over \partial r_i} - {2 \over 3} \, \delta_{ik}
\, {\partial v_l \over \partial r_l} \right)^2 ,
\label{enpro}
\end{equation}
where $\kappa$ is heat capacity coefficient and $\eta$ is
viscosity coefficient.  In the classical limit,
cf.~e.g.~\cite{dan84}, the coefficients are approximately given by
$\kappa \approx (75 \sqrt{\pi} /64 \sigma_{NN}) \sqrt{T/m}$ and
$\eta \approx (5 \sqrt{\pi} /16) \sqrt{mT} /\sigma_{NN}$.
However limited, an insight into the entropy production
during the expansion
may be obtained by substituting into (\ref{enpro}) results for
velocity, density, and~temperature from the~simplified
consideration in Sec.~\ref{expcen}, involving assumptions of
cylindrical symmetry,
self-similar expansion, and ideal-gas equation of state.
An examination of the r.h.s. of Eq. (\ref{enpro}) shows that
the heat conduction can
cause a drop of local entropy with time when temperatures
on the average decrease with moving away from a~given point,
whereas the viscosity
always leads to a rise in the local entropy per
particle. Substitution of
the results from Sec.~\ref{expcen} yields a net drop
of entropy at the center of the system with time,
quantitatively consistent with
what is observed in Fig.~\ref{profau}; temperatures decrease in
the surroundings due to the expansion into the vacuum.  The
substitution further yields
a~divergence of entropy production-rate towards the edges of
density distribution; in the net entropy production-rate the
viscous
contribution is integrable while the conduction contribution
is nonintegrable.  If a cut-off density is adopted for the
validity of hydrodynamics, then the net increase of entropy
per nucleon on the account of heat conduction in expansion,
may be estimated with
the formula
\begin{equation}
\delta \left( {S \over A} \right) \simeq {3 (\gamma -
1)^{1/2} \over \sigma_{NN} \, \rho
(0,0) \, R(0)} \left( \rho (0,0) \over \rho_f
\right)^{\gamma/2} , \label{risen}
\end{equation}
cf. Eq. (\ref{rhot}).  The increase due to viscosity turns out
to be given by a similar formula but with a numerical
coefficient
in front smaller by one order of magnitude.  Rise of the
entropy
production-rate towards the edges of density distribution is
quite consistent with what is observed in Fig.~\ref{profau}.
The results indicate that most entropy in expansion at
$b$~=~0 is produced
within an interface between the hot matter and the vacuum and
that the production is associated with the
equalization of temperature.  Quantitatively, Eq.~(\ref{risen})
gives an increase of the entropy during an expansion in the 400
MeV/nucleon reaction by
$\delta (S/A) \sim 1$, while the simulation yields $\delta
(S/A) \sim 0.6$.

Some limitation of the above consideration becomes apparent
when
a {\em self-consistent} solution of the viscous hydrodynamic
equations is confronted with simulations.  While
finding that the consistency in the solution moderates an
increase of the entropy,
Kapusta~\cite{kap81} obtained a~near independence of the
entropy
generated in expansion on the mass of an~expanding nuclear
system, $\delta (S/A) \sim 0.3$ for $\rho_1 \sim 2 \rho_0$.
By~contrast, in the simulations the generated entropy increases
when $A$ decreases, actually in nearly
such a~manner as predicted by Eq.~(\ref{risen}), $\delta (S/A)
\propto A^{-1/3}$.  The increase appears, in fact, consistent
with the central data for different
systems~\cite{cse87,dos88a}.

Following (\ref{sba}) or more detailed considerations,
generation
of entropy during expansion in the considered $b=0$ Au + Au
reaction, increases {\em local} kinetic energy towards
freeze-out
by $\sim 50\%$, limiting the rise of collective energy.
In principle, the dissipation could have been so strong that
no
collective energy would have been generated at all.  Then,
correspondingly, cluster yields would
have been lower also.  The correlation between cluster yields
and
collective motion is illustrated in Fig.~\ref{clus} which
shows the beam-energy dependence of yield ratios from standard
calculations and from calculations with the collective motion
extinguished by particle-position interchange, together with
data~\cite{dos88a}.  The beam-energy
dependence of the ratios of measured $A = 3$ cluster yields
to proton yield is
not well reproduced within the standard calculations; the
quality
of the description is about the same as within the QSM
model~\cite{dos88a}.  Unambiguously, though,
the data favor the results from the standard calculations with
a~significant collective motion developed, over the results
from calculations with the motion extinguished.

Dissipation heating the edges of matter distribution
and
cooling off the center of a~colliding system at
$b=0$ as in Fig.~\ref{profau}, beyond what actually can be seen
there,
together with some emission before equilibration, lead
to quite a spread of entropy values per baryon in the vicinity
of particle emission points, cf. Fig.~\ref{dsa}.  As,
generally, more clusters are emitted when entropy is low
than when it is high, the~average entropy in the vicinity of
cluster emission-points may be expected lower than in the
vicinity of nucleon emission-points.
Specifically, if the
distribution such as in Fig.~\ref{dsa} were approximated in a
gaussian form $d{\cal P}/d(S/A) \sim \mbox{exp} \left[- (S/A -
\overline{S/A})^2/2 \Delta^2(S/A)\right]$, and a complete
equilibrium were assumed with $M_X/M_N \sim
\mbox{exp}\left[-(A_X -1)(S/A)\right]$ for a given entropy, then the
average entropy in the
vicinity of emission points of baryons with different masses
would be expected to decrease with a~decrement per
unit mass $\langle S/A \rangle_{A+1} - \langle S/A
\rangle_{A}\;\,
\raisebox{-.5ex}{$\stackrel{<}{\scriptstyle\sim}$ } - \Delta(S/A)/2$.
In the 400~MeV/nucleon Au~+~Au reaction the dispersion is
$\Delta (S/A) \sim
0.40$, and the expected decrement would then be
$\langle S/A \rangle_{A+1} - \langle S/A
\rangle_{A}\;\,
\raisebox{-.5ex}{$\stackrel{<}{\scriptstyle\sim}$ } - 0.20$.
In~the~simulation, the
entropy in the vicinity of last collision points is found to
decrease by about 0.14 per mass unit for light baryons.
With an increase of the beam energy, the entropy generated in
expansion increases, just as does the entropy generated
within
shock fronts.  Likewise, the dispersion
of entropy values increases, and the mean entropy
in the vicinity of emission points is found to decrease faster
with baryon mass.
E.g. in the 1~GeV/nucleon $b=0$ Au~+~Au reaction-simulation the
decrease is about 0.20 per mass unit.  The~entropy
differences diminish, on the other hand, when the beam energy
is lowered (though, they become then increasingly more
difficult to assess reliably within the calculation).

In the analysis~\cite{kuh93}
of intermediate-mass fragment yields from central Au + Au
reactions, in particular at 400~MeV/nucleon, a~gradual decrease
of entropy with
charge was observed, however slower than in the present
calculation.
For~heavy fragments the decrease might be moderated by size
effects. In the
analysis~\cite{kuh93} the entropy values at the low-$Z$ end
were
further found lower than in~\cite{dos88a}, by~0.5 at
400~MeV/nucleon, which could become
clarified once masses get resolved and
angular coverage extended within that experiment.

At a finite $b$, generally, more variation of entropy values
throughout the system is expected than at $b = 0$, as
spectator regions
are characterized by $S/A \sim$ (1--2).  With regard to the
participant
region, a shock at a finite $b$ would lead to a lower entropy
than a shock at~$b = 0$, see Eq.~(\ref{ealpha}) and
Fig.~\ref{rankhu}.  If one assumed
that about
the same amount of entropy were produced in an~expansion at
a~finite $b$, as at $b=0$, then one would expect, on the basis
of the RH equation, a~change of entropy with $b$ to follow
\begin{equation}
\left( { S \over A} \right)_{b} - \left( { S \over A}
\right)_{b=0} = {S \over A} \left( \rho_1\left(0,E_{beam}'\right),E'
\right) - {S \over A} \left( \rho_1\left(0,E_{beam}\right),E \right) ,
\label{SAb}
\end{equation}
where entropy on the r.h.s.~is written as a function of
local baryon density and of net local energy per nucleon.  The
density~$\rho_1$ is from Eq.~(\ref{ranr}) and energies $E$ and
$E'$ are related, respectively, to $E_{beam}$ and $E_{beam}'$,
cf.~Eq.~(\ref{ealpha}), according to $E = m_N - 16\,\mbox{MeV} +
E_{beam}/2(1+ \sqrt{1+E_{beam}/2m_N})$.  The result for the
entropy as
a~function of $b$ from (\ref{SAb}) is shown in the right panel
of Fig.~\ref{enmu} with a dashed line.  The
decrease is clearly
in contradiction with the simulation and with data.  Note
that the experimental entropy~\cite{dos88a} is determined from
fragments
emitted into wide angles, i.e. stemming from the participant
region.  The~consideration, though, ignored till now the weak
discontinuity. As~the latter spreads, the~viscosity slows down
the nuclear matter, generating entropy.  If~one assumed that
kinetic energy were completely thermalized for a~given density,
then one would expect for the matter stemming from the center,
instead~of~(\ref{SAb}),
\begin{equation}
\left( { S \over A} \right)_{b} - \left( { S \over A}
\right)_{b=0} = {S \over A} \left( \rho_1\left(0,E_{beam}'\right),E
\right) - {S \over A} \left( \rho_1\left(0,E_{beam}\right),E \right) .
\label{Sab}
\end{equation}
This now leads to an increase of the entropy with an increase
of~$b$ as the thermalization for a~given energy occurs at
a~lower
density.  The increase is quadratic in $b/b_{max}$ for low
values of the reduced impact parameter, cf. (\ref{ranr}), in
a~qualitative agreement with the data.  The~difference in
entropy
between high and low $b$ increases with an increasing beam
energy, as seems to be the case with data~\cite{dos88a}.
Quantitatively, though, the entropy from~(\ref{Sab}) is still
too low compared to the large-multiplicity data
displayed in~Fig.~\ref{enmu} or the
simulation at large $b$.  (Although at a~maximum compression
the entropy gets, indeed, close to $(S/A)
\left( \rho_1\left(0,E_{beam}'\right),E \right)$.)
The result from~(\ref{Sab}) is shown
by a solid line in Fig.~\ref{enmu} and only the
difference between the solid and dashed lines can be attributed
to the viscosity.  The additional entropy at large $b$ is
associated with
the rise in the entropy generated during expansion, when the
size of
participant region diminishes.  For~$b$ close to maximum the
participant mass is~\cite{gos78} $A_{part} \simeq 3 \sqrt{2}\,
A \, (1 - b/b_{max})^2$, where $A\, (= 197)$ is the mass of one
nucleus. With the scaling in the simulations such as in
(\ref{risen}), the
generated entropy behaves then as $\delta (S/a) \propto (1 -
b/b_{max})^{2/3}$ for large $b$.

Summing up the discussion on entropy production, most entropy
at $b=0$ is due to shock waves and
some is generated in expansion.  As $b$ increases, the shock
contribution to the entropy in the participant region decreases,
and the
contribution of the spreading weak discontinuity rises~from~0.
Given that the density at the center gets close to $\rho_1$
from~Eq.~(\ref{ranr}) and entropy close to
$(S/A)
\left( \rho_1\left(0,E_{beam}'\right),E \right)$, it~is
apparent that properties of
matter in the compressed state at the center can be regulated
 by selecting impact parameters and
changing the beam
energy.
The entropy per nucleon
generated in expansion increases with~$b$.
Even at $b=0$ and a~high energy, the entropy in the vicinity of
particle
emission points is spread out around the mean entropy in
a~reaction, because of the dissipation.

\subsection{Collective Energy}
Very recently, values of mean baryon
energies~\cite{bar91,bau93,rei93,jeo94} and shapes of
spectra~\cite{rei93,jeo94,hsi94} have been used
 to
extract
the magnitude of collective energy in collisions.
(Procedures followed, in particular, Eq.~(\ref{avenre}) and
nonrelativistic versions~of Eqs.~(\ref{3d})~and~(\ref{dnX}), with
Coulomb corrections; see also~\cite{dan92}.)
It is then
worthwhile to discuss the experimental results on the energy
and compare them to calculations.

The collective-energy values of about 18, 31, and
51~MeV/nucleon,
deduced by the FOPI collaboration using intermediate mass
fragments~(IMF)~\cite{rei93} from central Au~+~Au reactions at
beam
energies of 150, 250, and 400~MeV/nucleon, come close to the
values of collective energy at freeze-out in present simulations,
of~21, 35, and 55~MeV/nucleon at the respective beam energies,
see~Fig.~\ref{colau}.  Revision~\cite{jeo94} of the result for
150~MeV/nucleon, with a lower bound on collective energy put
at~10~MeV/nucleon, stemmed from a~concern that studied events
were not exactly central, and the value of collective energy
was inflated due to a~persistance of longitudinal motion
and a particular angular range of the detector setup.  It~is
argued below that latter bound is likely too cautious.

As the separation of kinetic energy into collective and
thermal or excitation components occurs at freeze-out, this
separation should not depend very strongly on impact parameter
(cf. discussions in~Sec.~\ref{Central} and Ref.~\cite{dan92};
excitation energy is not expected to vary rapidly).
The collective energy per nucleon for light
baryons from simulations, as a function of impact parameter,
is~illustrated~in~Fig.~\ref{ecolb}.  Between
$b=0$ and $b=12$~fm the energy rises only~by~20\%.  For heavier
particles, and if statistical decays of spectator fragments
were taken into account, the rise would be generally faster.
Nonetheless, in~the~($p^{\perp}$-$y$) distribution presented
in~\cite{jeo94}
no traces of any spectator decay can be seen.  An~average
impact parameter is $\langle b \rangle \;
\raisebox{-.5ex}{$\stackrel{<}{\scriptstyle\sim}$ } 3$~fm, and
particle charge is $3 \le Z \le 8$.  Relative to $b = 0$, for
such charges and impact parameters, still only a~small
change of collective energy would be expected.  What definitely
changes with the impact parameter is the
division of
collective energy between the longitudinal and
transverse degrees of freedom.  If simulations adequately
described the situation in reactions, then, at
representative
impact parameters, the energy per transverse degree
of freedom
would be close to the longitudinal energy, and the angular
cuts
in~\cite{rei93,jeo94} {\em would not} affect the deduced value.
(Note, that the~equality of energies per degree of
freedom  does not contradict azimuthal
anisotropies~\cite{jeo94}, cf.~Sec.~\ref{expb}.)
Given an~unfavorable
situation for the determination of collective energy,
with an~anisotropy similar to that in the~simulation
at~$b=0$, but with more collective energy associated with the
longitudinal degree of freedom
than with one transverse, one could still put a~more
stringent lower bound on~the collective energy than
in~\cite{jeo94}, at~14~MeV/nucleon, taking into account
angular cuts and a~behavior of particle energies with~$Z$, and
assuming smooth momentum-distributions.
Analogous lower bounds, about 25\% below the~initially deduced
values~\cite{rei93}, at 27 and 40~MeV/nucleon, respectively,
could be put
on~the~collective energy in~the~central Au~+~Au collisions at
the two other beam energies of 250 and 400~MeV/nucleon.

At $b \sim 0$, or when examining transverse
directions~\cite{dan92},
one might expect a~gradual decrease of collective energy with
mass.  Thus, in~Fig.~\ref{profau} it can be seen
that regions with a high entropy, from which few
heavy particles would originate, freeze-out early and are
associated with relatively high collective velocities.  Regions
with low entropy, on the other hand, would freeze-out late and
would be associated with lower than average collective
velocities.  Spectra of~IMF, corresponding to wide
angles in the c.m., from~Au~+~Au reactions
at~100~MeV/nucleon and $b \;
\raisebox{-.5ex}{$\stackrel{<}{\scriptstyle\sim}$ } 4$ fm, were
analysed in~\cite{hsi94}, and collective energy per nucleon
gradually decreasing with fragment mass was deduced.
The~mean collective energy at the low mass end was
$\sim 9$~MeV/nucleon, which should be
compared~to~(cf.~\cite{dan92}) $(3/2)
E^{\perp}_{coll} \sim 11.5$~MeV/nucleon in the simulations
of that system at the respective impact
parameters.

\section{Dependence on Interactions}
\label{ints}

\subsection{Collective Energy}

Judging from Fig.~\ref{ecolb}, little sensitivity of the
collective energy to nuclear compressibility might be expected,
as the latter generally brings less change to nuclear
collisions than wide variations of impact parameter.
In fact, e.g.~in~the head-on 400~MeV/nucleon Au~+~Au reaction,
the net collective energy per nucleon is found the same,
within $\sim 1$~MeV, for $K = 200$~MeV and $K = 380$~MeV.
On~the~other hand, the~division of kinetic energy into
collective and excitation energy should be sensitive
to particle interaction cross-sections which directly affect
the freeze-out, see text below~Eq.~(\ref{Ecol}).
In the aforementioned Au~+~Au reaction, the~reduction of
interaction cross-sections throughout the system~by~30\%, and
the reduction following a~parametrization~\cite{kla93}
$\sigma_{NN}^{med} = \sigma_{NN} \exp(-\nu
\rho/\rho_0)$, with $\nu = 0.3$, lead to a drop in the
collective energy by~16\%,
and~10\%, respectively, compared to a~standard calculation.

On the other hand, a~sensitivity to
the nuclear compressibility can be expected for the
anisotropy
of collective energy which is associated with a delay
in the start of longitudinal expansion compared to
transverse.  The~delay time is
of the order of
\begin{equation}
t_{sh} = {2 R - W \over \gamma_0 (v_0 + v_1^{sh}) } = {(2R - W)
 (\rho_1 - \gamma_0 \rho_0) \over \gamma_0 v_0 \rho_1},
\end{equation}
where $W$ is shock width in terms of a~distance in normal
matter, introduced in~an~attempt
to account for the mean-free-path effects, and $v_1^{sh}$ is
shock speed with respect to compressed matter.  Relative to the
characteristic time scale for transverse expansion $t_{exp} =
R/c_s$ (characteristic generally~\cite{lan59,zel66},
beyond the consideration in~Sec.~\ref{expcen}),
where $c_s$ refers to compressed matter,
the~delay time is
\begin{equation}
{t_{sh} \over t_{exp}} = {(2 - W/R) \, c_s \over \gamma_0 (v_0
+ v_1^{sh}) }.
\label{tse}
\end{equation}
Small changes in the relative delay time lead to finite changes
in the energy components, affecting anisotropy.
At any one intermediate beam energy, the
ratio~(\ref{tse}) is typically
larger by $\sim 17\%$ for the stiff equation of state than for
the~soft, and, correspondingly, the~difference between
transverse energy per degree
of freedom and longitudinal energy is larger for the~stiff
equation than for the soft.
At
intermediate energies
the shock width in~(\ref{tse})
may be expected, on~the basis of
Navier-Stokes equations~\cite{dan84}, to be of a~magnitude
$W \simeq 2 \widetilde{\lambda}$, where
$\widetilde{\lambda} = 1/(\widetilde{\sigma} \rho_0)$, and where
$\widetilde{\sigma} \simeq 30$~mb
is an~isotropic cross-section that yields the same transport
properties as the differential NN cross-section~\cite{dan84}.
This roughly conforms
with $W \sim 5$~fm seen in~Fig.~\ref{profau}.  For~the stiff
equation of state in the 400~MeV/nucleon Au~+~Au collision, the
ratio~(\ref{tse}) is then estimated at about~1 ($c_s
\sim 0.6c$ for the state from RH equation), and for the soft
equation at about~0.85 ($c_s \sim 0.45c$).  The reduction in
the
relative delay time with the change of compressibility leads to
an~increase of
longitudinal energy at the cost of transverse energy by $\sim
2$~MeV/nucleon in the Au~+~Au simulation (see
Fig.~\ref{colau} for energy values), decreasing the
anisotropy of the {\em collective} energy at freeze-out,
$\chi_{coll} = ( E_{col}^{\perp } /2 -
E_{coll}^{\parallel} )/ E_{coll}^{\parallel} $, from
$\chi_{coll} \simeq 0.8$ to $\chi_{coll} \simeq 0.45$.

Following~(\ref{tse}), a reduction in cross sections (or, more
generally, interaction rates) compared to free space by $\sim
20\%$ may lead to a~similar decrease in the delay time and
in collective-energy anisotropy, as the change in the
compressibility above.  Indeed, in a simulation of the
Au~+~Au reaction using a~stiff equation of state, the
reduction of cross section by 20\% gives $\chi_{coll} \simeq
0.40$.
A~near-isotropy in collective energy could be expected,
following~(\ref{tse}),
for a~40-50\% reduction in cross sections.  In~a~simulation
the~collective energy components per degree of
freedom become identical when cross sections are
reduced by~40\%.

On the basis of (\ref{tse}) and more generally, one might expect
similar results from the dynamics when varying the size of
a~system by a factor, as when varying cross sections by
the same factor.  Results would need, in fact, to be the same,
up to a general rescaling of multiplicities, if
nuclei were
sharp-edged spheres.  (It would not matter whether r.h.s. of
Eqs.~(\ref{boltz}) and~(\ref{rboltz}) were multiplied by
some factor, or l.h.s. were divided by the same factor; see
also~\cite{bon88}.)
On the basis of the tests above, one might specifically
expect that an~isotropy of collective energy would be obtained
if nuclear radius were reduced by 40\%, i.e. mass reduced
to $A \sim 40$.  Simulation of a~head-on 400~MeV/nucleon
Ca~+~Ca reaction yields a~result in a~disagreement with this
expectation, $\chi_{coll} \simeq -0.25$, demonstrating the
importance of corona effects for a light system as compared to
heavy~\cite{sto86}, given a~similar
width of the nuclear surface region.

The corona effects play a lesser role when only directions
pointing away from
the beam axis are examined.  Consequently, scalings
expected on the
basis of~(\ref{tse}) may be followed to a~larger
extent at a~finite $b/b_{max}$, than at $b = 0$.
The~ratio~(\ref{tse}),
without any modifications, is relevant for
dynamics~at~$b/b_{max} \;\,
\raisebox{-.5ex}{$\stackrel{<}{\scriptstyle\sim}$ } 1/3$.  At
$b/b_{max} \sim 0.25$ and at the beam energy
of~400~MeV/nucleon, the anisotropy of
{\em particle} energies $\chi = \langle E^y \rangle / \langle
E^{x'} \rangle - 1$, generally somewhat
lower than $\chi_{coll}$, is $\chi
\simeq 0.65$ in~a~Au~+~Au simulation when using stiff
equation of
state, and $\chi \simeq$ 0.40, 0.45, and 0.40, respectively,
when using soft equation, when reducing collision rates by
20\%,
and when turning to a~Nb~+~Nb system with nuclei of
a~$\sim 20$\% lesser radius than Au.

The~400~MeV/nucleon data~\cite{gut90} displayed in
Fig.~\ref{rla}(b)
rule out the soft equation of state with momentum dependence in
the optical
potential missing, in a~combination with free-space or lesser
cross-sections.
After modifing the~elements of calculated kinetic-energy
tensor on the account of the fluctuations of estimated
reaction-plane direction in the Au~+~Au reaction, the
anisotropy of particle energies
for the soft equation of state would continue to be about
0.6 of the anistropy for the stiff equation as above.
A~corresponding maximum value of
$R_{\lambda} \sim 1.26$ for the soft equation, as a function of
the centrality, would fall
below the data.  On~the other hand, the data would allow for
the stiff equation of state in a~combination with cross
sections reduced by 10\% at the particular beam energy compared
to free space.

\subsection{Hollow Structures}

In transport simulations of head-on Mo~+~Mo collisions at 60
and 100~MeV/nucleon, Moretto~{\em et al.}~\cite{mor92} noted
a~formation
of disk structures perpendicular to beam axis, which fragmented
with time.  Further analyses~\cite{bau92,bao92} revealed
openings at disk centers.  In the simulations~\cite{bor92} of
central collisions
of heavy nuclei at 30~MeV/nucleon,
bubble structures were observed.
Hollows were attributed in~\cite{bau92} to a~rarefaction wave
reaching the center of a~system, while in~\cite{bor92} the
issues of Coulomb stability were stressed.  The~structures
generated some experimental interest.  Below, the mechanism of
the formation of hollow structures is critically reassessed;
the particular issues might not be of a~grave importance,
but the~different arguments are being repeated in the literature.

Despite of some increase in their width~\cite{dan84},
shock-like interfaces continue
to develop in the~simulations of central reactions of heavy
nuclei, down to beam energies per nucleon of few
tens of~MeV. Differences between maximum density in the
simulations and the~RH~expectations at the lower beam
energies in~Fig.~\ref{rankhu}, are partially
associated with a~Coulomb repulsion~\cite{dan92}.  Between
the shocks in reactions, an~expansion develops in the
directions parallel to shock surfaces, what gives
a~planarity to
transient structures.  At~energies less than~35~MeV/nucleon, or
for cross-sections reduced as~compared to free space, and/or
for lighter nuclei, the shock width becomes as large as the
nuclear diameter and
a~reacting system equilibrates then as a~whole.

The onset of transient structures is
marked in~Fig.~\ref{frezen} with some~increase of density in
the vicinity of last particle collisions, towards low beam
energies.
The~fact that structures persist in simulations up to the
energies in excess of~100~MeV/nucleon is necessarily
conditioned on
a~massive emission
of rapid particles early on in reactions (if few particles
were emitted, then structure formation would be limited to
the beam energies less than 4 times the binding energy,
i.e.~$\sim
30$~MeV/nucleon).  The~simulations of a~Mo~+~Mo
system
with  a~total mass of~196, within the present model, e.g. yield
residues of a~mass $\sim
56$, $\sim 105$, and $\sim 177$, at the
beam energies of 110,
60, and~20~MeV/nucleon, respectively.  Figure~\ref{profmo}
displays baryon
density, radial velocity, and entropy along and perpendicular
to the beam axis from the 60~MeV/nucleon $b = 0$ Mo~+~Mo
simulations
with and without Coulomb interactions.  Emission
is evidenced in extended density tails at times $t \;\,
\raisebox{-.5ex}{$\stackrel{>}{\scriptstyle\sim}$ } 50$ fm/c.

The~argumentation in~\cite{bau92} implicitly attributes hollow
formation to details in a~density distribution after
equilibration, and
specifically to a~sharp-edged
surface.  Following the arguments there, one~would find it
difficult to understand
why hollows do not develop within the simple solution
from~Sec.~\ref{expcen}, or~within solutions
in~Refs.~\cite{zel66,bon78}, or in reaction simulations at
$E_{beam} > 150$~MeV/nucleon, or, with reference to everyday
experience,
why rings do not form when spattering water on a~tabletop or
a~wall.
An~edge in the density distribution might be
built onto such a~solution as in~Sec.~\ref{expcen} as a~perturbation.
A~perturbation would satisfy a~classical wave equation.  With
time the initial pulse would move inwards at a~speed of sound
relative to remaining matter, and emerge from the
center as
a~wiggle in the case of a~cyllindrical symmetry, with a~peak
in front followed by a~dip, and as as a~dip
in the case of a~spherical
symmetry.
(Phase shift at the center for the Fourier components of the pulse
is $\pi/2$
in the case of a~cyllindrical symmetry and $\pi$ in the case of
a~spherical symmetry.)
As~can be seen in~Fig.~\ref{profmo}, the~expansion in
simulations develops past the~maximum in $dv /dr$ in
transverse direction, at $t \sim
36$~fm/c, without any~noticable depletion in the central
density relative to outer regions.  This is due to
a~continuous behavior of parameters of nuclear matter after
equilibration.  A~depletion appears later and is associated
with the changes in outer regions of an~expanding system.

As the~nuclear system expands at a~low entropy $S/A \sim 1$,
see~Fig.~\ref{profmo}, down to subnormal densities, a~negative
pressure develops, cf.~Fig.~\ref{eos}.
As~a result,
outer edges of the expanding matter begin
to deccelerate
(the vacuum pressure
is zero); see the plateau in transverse velocity
in~Fig.~\ref{profmo} developing
after central density falls below $0.9\,\rho_0$.
In~consequence,
with time the density becomes enhanced at the outer edges as
compared to the center.  As~the system is driven in the
meantime into a~region of adiabatic instability within
thermodynamic parameters (see also~\cite{dan79,ber88}),
characterized by $(\partial P / \partial \rho)_{S/A} < 0$,
the~pressure becomes more negative at the outskirts with
a~higher density, than at the center.  From the center then
the matter
continues to move out and accumulate in the outer regions.
The~system
is being cooled off all the time by emission.  Until $t \sim
70$~fm/c from the Mo~+~Mo system displayed
in~Fig.~\ref{profmo}, particles with a~total mass of~$\sim 35$
are emitted.  At time $t \sim 100$~fm/c the total mass of
emitted particles rises to $\sim 60$.
Apart from the pressure within the system, effects of which are
local, Coulomb interactions act.  With time, these render
a~stability to a~hollow structure that forms as the matter
contracts at the outskirts of the system.  When the Coulomb
interactions are missing or are weak, the matter having first
contracted
at the outskirts, gradually collapses onto the center,
see~Fig.~\ref{profmo}(b).  This is the case in simulations of
such light systems as~Ca~+~Ca.

Deficiency of the single-particle models is the lack of
fluctuations.  Because of the fluctuations, a~system might, in
reality, not condense into a~ring or a~bubble, but into
a~number
of clusters.  If~a~structure were formed, it would later break,
due to fluctuations, into clusters of which diameter would be
likely commensurate with the minor diameter of a~ring, or
the thickness of a~bubble.  In~any~of these cases a~large
number of~IMF would be produced, slow in the c.m.s. on a~scale
of collective velocities characterizing the early emission.
The~total mass of these IMF, e.g. at the beam energy above,
could near half of the
mass of the system.  The~unusual abundance of these fragments,
already indicated~in~\cite{hxu93}, might thus serve as
a~signature of the development of an~adiabatic instability
over
a~macroscopic region in reactions.  The~latter underlies both
of the scenarios outlined above.
Evidence
for a~planarity in the emission of the slow IMF would
additionally show that the instability sets in on
a~dynamic path for the system.  Search for signatures of
transient structures in light particles might, on the other
hand,
be hopeless.  These particles are primarily emitted when the
fate of matter left behind has not yet been decided.

\section{Conclusions}
\label{Conclu}

Dynamics of energetic symmetric heavy-ion reactions has
been analyzed
at a~qualitative level, with~an~attention directed at
the~collective
behavior of nuclear matter.  The~analysis relied on
transport-model reaction-simulations, analytic and
near-analytic considerations, and measurements.

In~the
simulations
of head-on reactions, nearly completely developed shock-fronts
may be identified, propagating into the projectile and target,
and separating the hot matter from normal.  In-between the
shocks, at finite impact parameters, a~tangential discontinuity
develops, that spreads with time.  Because of
small nuclear sizes and finite widths of the discontinuities,
they strongly
overlap.  Hot matter exposed to the vacuum in
sidewards directions begins to expand into these directions.
Expansion in the shock direction is delayed and does not
acquire same strength.  The~collective expansion affects
angular distributions, mean-energy components, shapes of
spectra and mean energies of different particles emitted into
any~one direction, and particle yields.

The stronger expansion perpendicular to the shock
direction leads to a~c.m.~90$^{\circ}$ peaking in the polar angle
in head-on collisions of heavy nuclei, and to a~peaking in the
azimuthal angle out of the reaction plane in semicentral
collisions.  These effects contrast with the naive expectations
regarding shocks, where one expects an~enhancement of the emission
in shock direction.  Unambigous experimental evidence
exists only for peaking of the emission in azimuthal angle.
Evidence for
a~motion of matter behind shocks, as characteristic for the
weak
discontinuity, exists in the strength of the sideward flow in
collisions.  The~strong overlap of the discontinuities dampens
this motion, nonetheless, to the extent that
spectator pieces can emerge in the high-energy collisions.

The maximum density in reactions falls somewhat below the
RH~expectations.  As~a~function of the impact parameter,
this~density follows well the scaling relation~(\ref{ranr}),
falling off quadratically with the impact parameter, while the
entropy at the center gets close to that
characteristic for stopped matter.  It~follows then that the
thermodynamic parameters of matter in the central region of
reactions can be regulated by changing the beam energy and
selecting impact parameter.  With regard to expansion,
its~relative strength
along and perpendicular to the shock front motion
depends on the magnitude of time necessary for a~shock
propagation through a~nucleus, relative to the time necessary
for the development of expansion.  The~relative delay time is
larger for the stiff equation of state than for the soft, and
the anisotropies in emission, as quantified e.g. using
collective-energy or particle-energy components, are larger for
the stiff equation of state than for the~soft.  The~delay time
depends, moreover, on the interaction cross sections that
determine the width of the shock front and the size of a~region
within which original equilibration occurs.  The~division of
the energy into collective and thermal for a given system
at~low~$b$, occuring at freeze-out, depends
solely on the cross sections.

The effects of expansion on particle spectra at any one c.m.
angle in central reactions, are the slopes decreasing with
increasing particle mass and downward concavities in baryon
spectra.  The effect on pion spectra is stronger than
following from naive nonrelativistic considerations.
A~pion spectrum is overall flatter than could be expected from latter
considerations and an~upward concavity can
develop.  Resonance decays may, however, fill up this
concavity.  With regard to pion yields, the~missing energy
considered by Stock {\em et al.} when analysing data, may be
identified with
the collective energy not available locally for pion
production. The~number
pions and deltas actually decreases within a~reacting system
towards freeze-out, as the collective motion with participant
region acquires strength. Overall, the effect of expansion
reduces pion yields
by $\sim 50$\% compared to the~situation without expansion.

Collective expansion is not fully isentropic towards
freeze-out, with produced entropy being associated with
the heat conduction.  Besides, in reactions the entropy is
produced
in shock fronts and within a~weak discontinuity on account of
the shear viscosity.  The~shock contribution to the net entropy
maximizes in head-on reactions.  In~low-energy collisions,
collective~expansion initiates
the~formation of hollow structures.
The~process further involves negative
pressure
and adiabatic instability, and finally Coulomb
interactions.

As may be apparent, a~wealth of physical phenomena is
associated with the collective behavior of matter in central collisions.
This is in stark contrast with early conclusions on
heavy-ion collisions based on limited inclusive data which,
dominated by peripheral collisions, appeared consistent with
a~featureless global equilibrium within the~participant region.

\acknowledgements
The author thanks for the hospitality extended to him at the
Institute for Theoretical Physics at Santa Barbara and at the
University of Giessen where some of this work was
carried out.
This
work was partially supported by the National Science Foundation
under  Grant No. PY-9403666.

\newpage

\begin{figure}
\caption{
Proton (solid lines) and neutron (dashed
lines) density
profiles from solving the TF equations, together with
the empirical charge density profiles (dotted lines)
\protect\cite{jag74} for $^{40}$Ca and $^{208}$Pb.
}
\label{profil}
\end{figure}

\begin{figure}
\caption{
Contour plots of baryon density in the reaction plane in Au +
Au collisions at 400 MeV/nucleon.  The displayed contour lines
are for the densities $\rho/\rho_0$ = 0.1, 0.5, 1, 1.5, and 2.
}
\label{contours}
\end{figure}

\begin{figure}
\caption{
Baryon density (top panels), radial velocity (center
panels), and entropy per baryon along (solid lines) and
perpendicular (dashed lines)
to the beam axis at different indicated times, in the $b = 0$
collision at a beam energy of 400 MeV/nucleon.
}
\label{profau}
\end{figure}

\begin{figure}
\caption{
Lines show the baryon density (top panel), temperature (center
panel), and total entropy
per baryon (bottom panel) expected behind a developed shock
front at $b=0$, as a function of beam energy,
from Eq. (\protect\ref{ranf}).  Filled circles
indicate the maximum
density from the simulations of $b=0$ Au + Au collisions at
different beam
energies.  Crosses indicate the maximum density, scaled by a
$\gamma$-factor according to Eq.~(\protect\ref{ranr}), from the
simulations of 1
GeV/nucleon Au + Au collisions at $b$ = 3, 6, 8, 9, 10, 11, and
12 fm, plotted against the equivalent beam energy following
from Eq.~(\protect\ref{ranr}).
}
\label{rankhu}
\end{figure}

\begin{figure}
\caption{
Time dependence of
the baryon density at $r=0$ (top panel) and the participant
transverse collective
energy per nucleon (bottom panel) in the $b=0$ Au + Au reaction
at 400 MeV/nucleon.
}
\label{densau}
\end{figure}

\begin{figure}
\caption{
Mean baryon densities in units of normal density in the
vicinity of last collisions of emitted
particles from the head-on Au + Au reactions (solid line), and
in the vicinity of collisions from
which the final light-clusters originate (dashed line), as a
function of beam energy.
}
\label{frezen}
\end{figure}

\begin{figure}
\caption{
Distribution  in the spherical angle of protons (circles),
deuterons (triangles),
tritons (inverted triangles), pions (diamonds), and pions
emitted within first 14 fm/c (stars), as
a function of polar angle from the simulation of a $b=0$ Au +
Au collision at 1 GeV/nucleon.  The solid line indicates the
distribution of tritons expected on the basis of the formula
(\protect\ref{dNdOm}).
 }
\label{polau}
\end{figure}

\begin{figure}
\caption{
Transverse and longitudinal components of the mean energies
of particles emitted from $b = 0$
Au + Au reactions, as a function
of beam energy.
 }
\label{eneau}
\end{figure}

\begin{figure}
\caption{
Transverse and longitudinal components of the mean collective
energy
per nucleon at the time of last collision of baryons
stemming from $b=0$ Au + Au reactions, as a function of
beam energy.
 }
\label{colau}
\end{figure}

\begin{figure}
\caption{
The initial discontinuity between the projectile and target
velocities breaks at a finite $b$ into
two shock
fronts propagating into the projectile $P$ and target $T$
(thick solid lines), and a weak tangential discontinuity
in-between (dotted line).  Shock-front position {\em within}
projectile at a later time is indicated with a dashed line.
}
\label{sketch}
\end{figure}

\begin{figure}
\caption{
Baryon density $\rho$ (top panel),
normal and tangential
velocity-components $v_n$ and $v_t$ to the
discontinuities in nuclear-matter (center panel),
coinciding with
radial and polar
velocity-components,
and entropy per baryon (bottom panel) as a function of the
distance from the center of a 400 MeV/nucleon Au + Au system at
$b = 7$ fm, along the normal to the discontinuities.  At the
given time $t = 13.5$ fm/c the discontinuities (at their
centers) are inclined
at an angle $\alpha \sim 52^{\circ}$ relative to the beam axis.
 }
\label{tang}
\end{figure}

\begin{figure}
\caption{
Mean energy component out of the reaction plane at $y =0$ (top
panel), and at all rapidities (center panel), and a lower of
the eigenvalues of a relativistic tensor $\langle s^{ij}
\rangle$, associated with a direction within the reaction plane
(bottom panel), for protons (circles), deuterons (squares),
tritons (triangles), and helions (diamonds), emitted from 400
MeV/nucleon Au + Au reactions, as a function of the impact
parameter.  }
\label{enrec}
\end{figure}

\begin{figure}
\caption{
Mean transverse momentum component within the reaction plane,
divided by mass, as a function of rapidity for protons
(circles), deuterons (triangles), tritons (inverted
triangles), and pions (diamonds) from 1 GeV/nucleon Au + Au
collisions at $b$ = 3 fm (top panel) and $b$ = 7 fm (bottom)
panel.  For collisions at $b$ = 7 fm, also the mean momentum
for pions emitted within first 16 fm/c is shown (stars).
 }
\label{pxa}
\end{figure}

\begin{figure}
\caption{
Left panel shows the flow parameter (\protect\ref{four})
in Au + Au
reactions at 250 and 400 MeV/nucleons, from the measurements of
Ref.~\protect\cite{gus88}, as a function of the reduced
participant
proton multiplicity.  Right panel shows the flow parameter from
the simulations of Au + Au reactions at 400 MeV/nucleon, as a
function of impact parameter.
 }
\label{slope}
\end{figure}

\begin{figure}
\caption{
Parameters associated with the kinetic-energy tensor in
symmetric reactions at 400 MeV/nucleon. (a)  Flow angle from
the
calculated mean kinetic-energy tensor, as a function of reduced
impact parameter in Au~+~Au (stars) and Ca~+~Ca (diamonds)
reactions. (b) Ratio of the out-of-reaction-plane
matrix-element to the lower in-plane eigenvalue (crosses),
as a function of the normalized participant proton
multiplicity in Au~+~Au reactions,
from measurements of Ref.~\protect\cite{gut90},
compared to
the ratio of eigenvalues (stars) from a calculated tensor with
matrix
elements modified on account of fluctuations of the estimated
reaction plane about the true plane.
 }
\label{rla}
\end{figure}

\begin{figure}
\caption{
Mean kinetic energy divided by temperature in thermal
equilibrium, as a function of temperature divided by mass. }
\label{ET}
\end{figure}

\begin{figure}
\caption{
Baryon density along (solid lines) and
perpendicular (dashed lines)
to the beam axis at different indicated times, in the $b = 0$
La + La collision at a beam energy of 800 MeV/nucleon.
}
\label{profla}
\end{figure}

\begin{figure}
\caption{
Time dependence of
baryon density at $r=0$ (top panel), participant
collective
energy per nucleon (center panel), and total entropy in the
$b=0$ La + La reaction at 800 MeV/nucleon from the standard
calculation (solid line) and from the calculation with an
interchange of particle positions (short-dashed lines). }
\label{densla}
\end{figure}

\begin{figure}
\caption{
Time dependence of elastic NN collision-rate
(top panel),
rates of $\Delta$ production in NN
collisions and of absorption in $\Delta$N collisions
(center panel), and rates of $\Delta$ formation in
$\pi$N collisions and of $\Delta$ decay (bottom panel). }
\label{rates}
\end{figure}

\begin{figure}
\caption{
Time dependence of the number of pions and deltas
normalized using the
baryon
number, in the $b=0$ La+La reaction at 800 MeV/nucleon, from
a
standard calculation (solid line) and from a calculation with
an interchange of particle positions (short-dashed line).
The two horizontal long-dashed lines show the equilibrium
number of pions and deltas for a baryon density $\rho = 2.25
\rho_0$ and an energy per baryon of
1133
MeV and 1103 MeV, respectively.  Dots show the number obtained
by integrating
over space the local equilibrium values of pion and delta
density in a standard calculation.
 }
\label{deltap}
\end{figure}

\begin{figure}
\caption{
Isentropes in the plane of the ratio of pion-and-delta
density
to baryon density vs the baryon density (top panel), and in
the plane of the temperature vs the baryon density (bottom
panel).  Approximate locations of the 800 MeV/nucleon $b=0$
La+La system at $t=14\,$fm/c in the two planes are marked with
crosses.  Dashed lines show a portion of a chemical equilibrium
path with regard to the pion and delta number.
 }
\label{entro}
\end{figure}

\begin{figure}
\caption{
Momentum distribution of protons (circles), deuterons
(triangles), helions (inverted triangles), negative
(squares) and neutral (diamonds) pions
from central 800
MeV/nucleon La + La reaction,
 in the vicinity of 90$^{\circ}$ in the c.m.
Left panel shows the results of calculations at $b = 1$ fm
(filled symbols) and data of Ref.~\protect\cite{hay88} (open
symbols).
Right panel shows the results of calculations with particle
positions interchanged during evolution.
Solid lines in the left panel indicate results of the
instantaneous freeze-out model, for baryons and neutral pions.
Long-- and short--dashed lines for neutral pions indicate
a contribution of free pions at freeze-out and a contribution
from
$\Delta$ decays, respectively.  Straight parallel lines in the
right panel serve to guide the eye.
 }
\label{spectra}
\end{figure}

\begin{figure}
\caption
{Ratio of the mean pion multiplicity to
the number of participant nucleons in central symmetric
reactions, as a function of laboratory
energy.  The solid and dashed
lines represent, respectively, the results of standard
calculations and of calculations with particle positions
interchanged, for the La+La $b=0$ reactions.  The
circles represent data of Ref.~\protect\cite{sto82}.
 }
\label{pimul}
\end{figure}

\begin{figure}
\caption
{Distribution of collective velocities for baryons from 800
MeV/nucleon La + La reaction at $b = 1$ fm, calculated in the
vicinity of last collisions, within an angular range of
velocities $\theta = 60-120^{\circ}$.
 }
\label{coldis}
\end{figure}

\begin{figure}
\caption
{Negative-pion distribution in the vicinity of c.m.
90$^{\circ}$ in the central 1.8 GeV/nucleon Ar + KCl reaction.
Insets show pion asymmetry-coefficient defined in Eq.
(\protect\ref{asym}).  Open circles in  both panels indicate
data of Ref.~\protect\cite{bro84}; distributions are normalized
using event cross-section given there.  Filled circles in the
left panel indicate results of the present model at $b = 1.7$
fm.  Squares and
diamonds in the right panel indicate, respectively, the results
obtained using the cascade code \protect\cite{cug82} with
anisotropic and isotropic $N + N \leftrightarrow N + \Delta $
cross-sections.  It may be mentioned that the present model
yields results indistinguishable from those from the
cascade code \protect\cite{cug82}, when optical potential is
switched off and same assumptions on scattering are adopted.
 } \label{ar18}
\end{figure}

\begin{figure}
\caption
{
Distribution of neutral pions with c.m. momenta $p> 200$ MeV/c
(top panel) and protons (bottom panel) in the azimuthal angle
$\phi'$ with respect to the reaction plane, at a momentum
component within the reaction plane $p^{z'} = 0$.  The angle
$\Theta$ is the angle of rotation of the $p^{z'}$-axis from
the $p^z$-axis in the direction of $p^x$.
 } \label{dNfi}
\end{figure}

\begin{figure}
\caption{
Contour lines (solid) for baryon density in the reaction plane
in 1 GeV/nucleon Au + Au collision at $b = 7$ fm and $t = 15$
fm/c, with overlaid contour lines (dashed) for pion and delta
density.
Baryon contour lines are for density values $\rho/\rho_0$ = 0.1,
0.5, 1, 1.5, and 2.  Pion and delta lines are for density
values $(\rho_{\pi} + \rho_{\Delta})/\rho_{0}$ = 0.02, 0.1, and
0.2. }
\label{denspi}
\end{figure}

\begin{figure}
\caption{
Left panel shows entropy per nucleon as a function of reduced
participant proton multiplicity, determined from
400~MeV/nucleon
Au~+~Au wide-angle data of Ref.~\protect\cite{dos88a}.
Right panel shows, as a function of impact parameter, the
entropy per nucleon associated with nucleons emitted into wide
angles from reaction simulations (circles),
cf.~\protect\cite{dan91}, and predictions for the entropy from
Eq.~(\protect\ref{SAb})~(dashed line)
and from Eq.~(\protect\ref{Sab})~(solid line).
 }
\label{enmu}
\end{figure}

\begin{figure}
\caption
{Ratios of cluster yields to proton yields at wide angles, cf.
\protect\cite{dan91}, as a function of bombarding energy in the
head-on Au+Au reactions. The solid and dashed
lines represent, respectively, the results from standard
calculations and from calculations with particle positions
interchanged.  The dots
represent data of Ref. \protect\cite{dos88a}.
 }
\label{clus}
\end{figure}

\begin{figure}
\caption
{Distribution of entropy per baryon in the vicinity of
last collision points in the 400 MeV/nucleon $b = 0$ Au + Au
reaction simulation.
 }
\label{dsa}
\end{figure}

\begin{figure}
\caption
{Impact-parameter dependence of the mean participant collective
energy per nucleon (stars) and transverse collective energy per
nucleon (diamonds), calculated as in~Eq.~(\protect\ref{Ecol}),
in Au~+~Au reactions at 400 MeV/nucleon.  }
\label{ecolb}
\end{figure}

\begin{figure}
\caption{
Baryon density (top panels), radial velocity (center
panels), and entropy per baryon along (solid lines) and
perpendicular (dashed lines)
to the beam axis at different indicated times, in the head-on
Mo~+~Mo reaction at 60 MeV/nucleon, from a simulation with (a)
and without (b) Coulomb interactions.
}
\label{profmo}
\end{figure}

\begin{figure}
\caption{
Isentropes in the plane of pressure vs density for the stiff
equation of state.
 }
\label{eos}
\end{figure}


\end{document}